\documentclass[12pt]{article}
\usepackage{amsmath} 
\usepackage{longtable}
\usepackage{graphicx,epsfig}    
\usepackage{color}
\usepackage{cite}
\usepackage{url}
\usepackage{amssymb}
\usepackage{xspace}

\newcommand\as{\alpha_{\mathrm{S}}} 
\newcommand\f[2]{\frac{#1}{#2}}

\def\beq{\begin{equation}} 
\def\eeq{\end{equation}} 
\def\to{\rightarrow} 
\def\nn{\nonumber}

\def\b0{\beta_0}

\def\GE{\gamma_E}
\def\beeq{\begin{eqnarray}}
\def\eeeq{\end{eqnarray}}

\def\mur{\mu_R} 
\def\muf{\mu_F}
\def\mur2{\mu_R^2} 
\def\muf2{\mu_F^2}

\def\to{\rightarrow}
 
\def\nn{\nonumber}

\def\Mhh{M_{hh}}

\newcommand{\sqrtS}{\ensuremath{\sqrt{s}}}

\newcommand\Matrix{{\sc Matrix}\xspace}

\newcommand{\nnloFT}{NNLO$_{\mathrm{FTa}}$\xspace}
\newcommand{\nnloFTi}{NNLO$_{\mathrm{FTa-i}}$\xspace}
\newcommand{\nnllFTi}{NNLL$_{\mathrm{FTa-i}}$\xspace}

\usepackage[overlay,absolute]{textpos}
\newcommand\PlaceText[3]{%
\begin{textblock*}{10in}(#1,#2)  
#3
\end{textblock*}
}%

\newcommand {\apgt} {\ {\raise-.5ex\hbox{$\buildrel>\over\sim$}}\ }
\newcommand {\aplt} {\ {\raise-.5ex\hbox{$\buildrel<\over\sim$}}\ }

\newcommand\Tstrut{\rule{0pt}{3.0ex}}         
\newcommand\Bstrut{\rule[-1.5ex]{0pt}{0pt}}   

\title{Soft gluon resummation for Higgs boson pair \\[1ex]
 production including finite $M_t$ effects}

\author{Daniel de Florian$^1$ and Javier Mazzitelli$^2$
\\\\
\footnotesize{$^1$International Center for Advanced Studies (ICAS), ECyT-UNSAM,}
\\
\footnotesize{Campus Miguelete, 25 de Mayo y Francia, (1650) Buenos Aires, Argentina}
\\[1ex]
\footnotesize{$^2$Physik-Institut, Universit\"at Z\"urich, Winterthurerstrasse 190, CH-8057 Z\"urich, Switzerland}
}

\date{}

\begin{document}
\renewcommand{\thefootnote}{\fnsymbol{footnote}}
\pagenumbering{gobble}

\PlaceText{160mm}{10mm}{
\noindent \small
ZU-TH 26/18 \\
ICAS  35/18
}

\maketitle

\begin{abstract}
\noindent 

We perform the all orders resummation of threshold enhanced contributions for the Higgs boson pair production cross section via gluon fusion, including finite top quark mass ($M_t$) effects.
We present results for the total cross section and Higgs pair invariant mass ($M_{hh}$) distribution.
We obtain results at next-to-leading logarithmic accuracy (NLL) which retain the full $M_t$ dependence, and are matched to the full next-to-leading order (NLO) prediction. Our NLL+NLO results represent the most advanced prediction with full $M_t$ dependence for this process, and produce an increase of about $4\%$ in the total cross section with respect to the NLO result for LHC energies, and for a central scale $\mu_0 = M_{hh}/2$.
We also consistently combine the full NLL with the next-to-next-to-leading logarithmically (NNLL) accurate resummation computed in the Born-improved large-$M_t$ limit, and match it to the next-to-next-to-leading order approximation of Ref.~\cite{Grazzini:2018bsd}, so called \nnloFT.
We find that the resummation effects are very small at NNLL for $\mu_0 = M_{hh}/2$, in particular below $1\%$ at 13~TeV, indicating that the perturbative expansion is under control.
In all cases the resummation effects are found to be substantially larger for the central scale $\mu_0 = M_{hh}$, resulting in a more stable cross section with respect to scale variations than the fixed order calculation.

\end{abstract}

\newpage

\pagenumbering{arabic}
\section{Introduction}

The study of the properties of the Higgs boson discovered by the ATLAS and CMS collaborations is one of the main goals of the present and future runs of the LHC.
Among the different measurements that can help to distinguish between the Standard Model (SM) and new physics scenarios, the measurement of the Higgs self coupling is one of particular interest, as in the SM it is determined by the scalar potential, responsible for the electroweak symmetry breaking mechanism.

The production of Higgs boson pairs provides a direct way of measuring the Higgs trilinear coupling, and the high-luminosity upgrade of the LHC is expected to provide constraints on its value by measuring the double Higgs production cross section \cite{CMS-PAS-FTR-16-002,ATL-PHYS-PUB-2017-001}.
In the SM, the main production mechanism is the fusion of gluons via a heavy quark (mainly top quark) loop, and the corresponding cross section has been computed at leading order (LO) in Refs.~\cite{Eboli:1987dy,Glover:1987nx,Plehn:1996wb}.
The QCD corrections for this process have been computed first in the heavy top-quark mass ($M_t$) limit (HTL), both at next-to-leading order \cite{Dawson:1998py} (NLO) and next-to-next-to-leading order \cite{deFlorian:2013uza,deFlorian:2013jea,Grigo:2014jma,deFlorian:2016uhr} (NNLO), and more recently the NLO corrections with full $M_t$ dependence also became available \cite{Borowka:2016ehy,Borowka:2016ypz}, later also supplemented by transverse momentum resummation \cite{Ferrera:2016prr} and parton shower effects \cite{Heinrich:2017kxx,Jones:2017giv}.
The size of the QCD corrections was found to be large --about a $70\%$ increase in the total cross section at NLO for LHC energies--, and also the difference with respect to the HTL was found to be significant, the latter being around $15\%$ larger than the full NLO result at 14~TeV.

Very recently, an improved and fully differential NNLO prediction --labeled \nnloFT for full-theory approximation, see also Refs.~\cite{Frederix:2014hta,Maltoni:2014eza}-- was presented in Ref.~\cite{Grazzini:2018bsd}, which in particular features the full loop-induced double-real corrections.
This result predicts an additional increase in the total cross section with respect to the full NLO calculation of about $12\%$ at the LHC, and a residual uncertainty due to missing finite-$M_t$ effects estimated to be about $2.5\%$.

Besides the previously described fixed-order calculations, the all-orders resummation of soft gluon emissions has also been performed --again within the HTL-- at next-to-next-to-leading logarithmic accuracy (NNLL) in Refs.~\cite{Shao:2013bz,deFlorian:2015moa}. The resummed contributions, which account for the dominant effect of the  missing higher-orders in the perturbative expansion in the threshold limit, are found to further stabilize the cross section leading to smaller theoretical uncertainties.

In this work we perform the resummation of the threshold enhanced contributions including finite $M_t$ effects.
In particular, up to next-to-leading logarithmic accuracy (NLL) we retain the full $M_t$ dependence, therefore obtaining NLL+NLO results that represent the most advanced prediction computed in the full theory.
Finally, by performing matching  to the \nnloFT cross section, we achieve {\it the state of the art } results for Higgs pair production by reaching NNLL accuracy within the best available approximation for the $M_t$ effects .

This work is organized as follows: in section~\ref{sec:resum} we collect all the analytical expressions needed to perform threshold resummation up to NNLL, then in section~\ref{sec:results} we present our numerical predictions for the LHC and future colliders, and  in section~\ref{sec:conclusions} we summarize the results.

\section{Threshold resummation}
\label{sec:resum}

We consider the hadronic production of Higgs boson pairs via gluon fusion.
The hadronic cross section for a collider center-of-mass energy $s_H$,  differential in the Higgs pair system invariant mass $\Mhh$, can be expressed in the following way
\begin{align}
\label{had}
\Mhh^2\,\f{d\sigma}{d\Mhh^2}(s_H,\Mhh^2) \equiv \sigma(\tau,\Mhh^2) =& 
\sum_{a,b} \int_0^1 dx_1 \;dx_2 \; f_{a/h_1}(x_1,\mu_F^2) 
\;f_{b/h_2}(x_2,\mu_F^2) 
 \\
& \hspace*{-1cm} \times
\int_0^1 dz \;\delta\!\left(z -
\frac{\tau}{x_1x_2}\right) 
 \hat{\sigma}_0\,z\;G_{ab}(z;\as(\mu_R^2), \Mhh^2/\mu_R^2;\Mhh^2/\mu_F^2) \;,\nn
\end{align}
where $\tau=\Mhh^2/s_H$, $\mu_R$ and $\mu_F$ are the renormalization and factorization scales respectively, and $\hat{\sigma}_0$ represents the Born level partonic cross section. 
The parton densities of the colliding hadrons are denoted by 
$f_{a/h}(x,\mu_F^2)$ with the subscripts $a,b$ labeling the type
of massless partons ($a,b=g,q_f,{\bar q}_f$,
with $N_f=5$ different flavours of light quarks).
The hard coefficient function $G_{ab}$ can be computed in perturbation theory, expanding it in terms of powers of the ($\overline{\text{MS}}$ renormalized) QCD coupling $\as(\mu_R^2)$ as:
\begin{align}
\label{expansion}
G_{ab}(z;\as, \Mhh^2/\mu_R^2;\Mhh^2/\mu_F^2) &=
\sum_{n=0}^{+\infty} \left(\f{\as}{2\pi}\right)^n
\;G_{ab}^{(n)}(z;\Mhh^2/\mu_R^2;\Mhh^2/\mu_F^2)\;.
\end{align}

We introduce now the notation needed to perform the soft gluon resummation in Mellin space \cite{Sterman:1986aj,Catani:1989ne}.
We start by considering the Mellin transform of the hadronic cross section, 
\beq
\label{sigman}
\sigma_N(\Mhh^2) \equiv \int_0^1 \;d\tau \;\tau^{N-1} \;
\sigma (\tau,\Mhh^2) 
\,,
\eeq 
which takes the following factorized form
\begin{equation}
\label{hadn}
\sigma_{N-1}(\Mhh^2) = \hat{\sigma}_0 \;\sum_{a,b}
\; f_{a/h_1, \, N}(\mu_F^2) \; f_{b/h_2, \, N}(\mu_F^2) 
\; {G}_{ab,\, N}(\as, \Mhh^2/\mu_R^2;\Mhh^2/\mu_F^2) \;.
\end{equation}
Here we have introduced the $N$-moments of the hard coefficient function and parton distributions, specifically
\begin{align} 
\label{pdfn}
f_{a/h, \, N}(\mu_F^2) &= \int_0^1 \;dx \;x^{N-1} \;
f_{a/h}(x,\mu_F^2) \;, \\
\label{gndef}
G_{ab,\, N} &= \int_0^1 dz \;z^{N-1} \;G_{ab}(z) \;\;.
\end{align}
Once all the ingredients in $N$-space are known, we can obtain the physical cross section via Mellin inversion,
\begin{align}
\sigma (\tau,\Mhh^2)
= \hat{\sigma}_0 \;\sum_{a,b} & 
\;\int_{C_{MP}-i\infty}^{C_{MP}+i\infty}
\;\frac{dN}{2\pi i} \; \tau^{-N+1} \;
f_{a/h_1, \, N}(\mu_F^2) \; f_{b/h_2, \, N}(\mu_F^2) \nonumber \\
\label{invmt}
& \times
\; {G}_{ab,\, N}(\as, \Mhh^2/\mu_R^2;\Mhh^2/\mu_F^2) \;,
\end{align} 
where the constant $C_{MP}$ defining the integration contour in the $N$-plane is on the right of all the possible singularities of the integrand \cite{Catani:1996yz}.

We will perform the all-order summation of the threshold enhanced  contributions, which corresponds to the limit $z \to 1$ or equivalently $N\to\infty$ in Mellin space, and appear as $\as^n \ln^m N$ terms with $1 \leq m \leq 2n$.
We will therefore consider (for the resummed contributions) only the gluon-initiated configuration, given that it is the only partonic channel that is not suppressed  in this limit.
The soft-gluon contributions in the large-$N$ limit can be organized in the following all-order resummation formula for the partonic coefficient function in Mellin space,
\begin{align}
\label{resfdelta}
G_{{gg},\, N}^{{\rm (res)}}(\as, \Mhh^2/\mu_R^2;\Mhh^2/\mu_F^2) 
&=  
C_{gg}(\as,\Mhh^2/\mu^2_R;\Mhh^2/\mu_F^2) \nn \\ 
&\cdot  \Delta_{N}(\as,\Mhh^2/\mu^2_R;\Mhh^2/\mu_F^2) +
{\cal O}(1/N)\; .
\end{align}
All the large logarithmic corrections are exponentiated in the Sudakov factor $\Delta_N$, only depending on the dynamics of soft gluon emissions from the initial state partons.
It can be expanded as
\begin{align} 
\label{calgnnll} 
~\vspace{-.5cm} \ln \Delta_{N}  \!\left(\as,\ln N;\frac{\Mhh^2}{\mu^2_R}, 
\frac{\Mhh^2}{\mu_F^2}\right) &=
\ln N \; g^{(1)}(\beta_0 \as \ln N) + 
g^{(2)}(\beta_0 \as \ln N, \Mhh^2/\mu^2_R;\Mhh^2/\mu_F^2 )
\nonumber 
\\ 
&+ \as 
\;g^{(3)}(\beta_0 \as\ln N,\Mhh^2/\mu^2_R;\Mhh^2/\mu_F^2 ) 
\nonumber \\
&+ \sum_{n=4}^{+\infty}  \as^{n-2}
\; g^{(n)}(\beta_0 \as\ln N,\Mhh^2/\mu^2_R;\Mhh^2/\mu_F^2 )\;. 
\end{align} 
The term
$\ln N \; g^{(1)}$ resums all the LL contributions
$\as^n \ln^{n+1}N$, $g^{(2)}$ collects the NLL terms $\as^n \ln^{n}N$, $\as\, g^{(3)}$ contains the NNLL terms 
$\as^{n+1} \ln^{n}N$, and so forth.
The perturbative coefficients $g^{(n)}$ needed to perform NNLL resummation are known and only depend on the type of incoming partons, and their explicit expression can be found, for instance, in Refs.~\cite{Catani:2003zt,Vogt:2000ci}.

All the contributions that are constant in the large-$N$ limit are contained in the function $C_{gg}(\as)$. They originate in non-logarithmic soft contributions and hard virtual corrections, and can be expanded in powers of the strong coupling:
\begin{equation}
\label{Cfun}
C_{gg}(\as,\Mhh^2/\mu^2_R;\Mhh^2/\mu_F^2) =  
1 + \sum_{n=1}^{+\infty} \;  
\left( \frac{\as}{2\pi} \right)^n \; 
C_{gg}^{(n)}(\Mhh^2/\mu^2_R;\Mhh^2/\mu_F^2) \;\;.
\end{equation}
In particular, in order to perform N$^i$LL resummation we need up to the $C_{gg}^{(i)}$ coefficient.
At the same time, this coefficient can be obtained from the N$^i$LO fixed order computation; even more, given that the soft gluon contributions  in $C_{gg}^{(i)}$ are universal, the only process dependence enters via the virtual corrections.
The explicit (universal) relation between $C_{gg}^{(i)}$ and the loop corrections has been derived up to $i = 2$ in Ref.~\cite{deFlorian:2012za}, and later at one order higher in Ref.~\cite{Catani:2014uta}, and reads (for $\mu_R = \mu_F = M_{hh}$)
\beeq
C_{gg}^{(1)}&=&
C_A \f{4\pi^2}{3} + 4 C_A \GE^2 + \frac{\hat{\sigma}^{(1)}_{\text{fin}}}{\hat{\sigma}_0}\;,
\\
C_{gg}^{(2)}&=&
C_A^2 \bigg(
-\frac{55 \zeta_3}{36}-14 \gamma_E  \zeta_3+\frac{607}{81}+\frac{404 \gamma_E
   }{27}+\frac{134 \gamma_E ^2}{9}+\frac{44 \gamma_E ^3}{9}+8\GE^4 \nn\\
   &+&\frac{67 \pi
   ^2}{16}+\frac{14 \gamma_E ^2 \pi ^2}{3}+\frac{91 \pi ^4}{144}
\bigg)
+C_A N_f \left(
\frac{5 \zeta_3}{18}-\frac{82}{81}-\frac{56 \gamma_E }{27}-\frac{20 \gamma_E
   ^2}{9}-\frac{8 \gamma_E ^3}{9}-\frac{5 \pi ^2}{8}
\right)\nn\\
&+& \beta_0^2 \f{11\pi^4}{3}
+C_A \frac{\hat{\sigma}^{(1)}_{\text{fin}}}{\hat{\sigma}_0}
\left(\f{4\pi^2}{3}+4\GE^2\right)+
\frac{\hat{\sigma}^{(2)}_{\text{fin}}}{\hat{\sigma}_0}\;,
\eeeq
where $\zeta_n$ represents the Riemann zeta function, $\GE$ is the Euler number and $\beta_0=(11C_A-2N_f)/12\pi$.
The infrared-regulated one and two-loop corrections $\hat\sigma^{(1)}_{\text{fin}}$ and $\hat\sigma^{(2)}_{\text{fin}}$ can be obtained from the corresponding matrix elements after applying the corresponding subtraction operator. The explicit formulas can be found in Ref.~\cite{deFlorian:2012za}.
For the particular case of Higgs boson pair production, their explicit expression valid in the HTL can be found in Ref.~\cite{deFlorian:2015moa}, while for the NLL resummation with full $M_t$ dependence we can obtain numerical results for $\hat{\sigma}^{(1)}_\text{fin}$, and therefore $C_{gg}^{(1)}$, using the publicly available grid interpolation of the two-loop NLO virtual corrections \cite{Heinrich:2017kxx}.

Finally, in order to fully profit from the knowledge of the fixed order calculation, we implement the corresponding matching.
As usual, we expand the resummed N$^i$LL cross section to ${\cal O}(\alpha_s^i)$\footnote{Relative to the LO $\as^2$ power, which is always understood.},  add the full N$^i$LO cross section, and subtract the expanded result of the resummed one to avoid a double counting of logarithmic fixed order effects, as
\beq
\sigma^\text{N$^i$LL+N$^i$LO} (s_H,Q^2) = \sigma^\text{N$^i$LL}_\text{res} (s_H,Q^2) - \sigma^\text{N$^i$LL}_\text{res} (s_H,Q^2) \Big |_{{\cal O}(\alpha_s^i)} + \sigma^\text{N$^i$LO} (s_H,Q^2) \, .
\eeq

\section{Numerical results}
\label{sec:results}

In this section we present the numerical predictions for the LHC and future hadron colliders.
We use the values $M_h = 125\text{ GeV}$ and $M_t = 173\text{ GeV}$ for the Higgs boson and top quark masses, and do not consider bottom quark contributions.
We use the PDF4LHC15 sets~\cite{Butterworth:2015oua,Ball:2014uwa,Dulat:2015mca,Harland-Lang:2014zoa,Gao:2013bia,Carrazza:2015aoa} for the parton densities and strong coupling, evaluated at each corresponding perturbative order.
The fixed order cross sections are obtained from the implementation of Ref.~\cite{Grazzini:2018bsd}, which is based on the publicly available computational framework \Matrix \cite{Grazzini:2017mhc}.

In the first place, we present in section \ref{sec:nll} the NLL+NLO predictions.
It is worth to point out that, even if more advanced predictions have been obtained for this process (specifically the so-called \nnloFT defined in Ref.~\cite{Grazzini:2018bsd}), these results represent the most advanced prediction computed in the full theory, i.e. with full $M_t$ dependence.

Based on the knowledge of the threshold enhanced contributions at NLL with full $M_t$ dependence, and in particular on the ${\cal O}(\as^2)$ of its expansion, we can also provide an improved fixed order (approximated) NNLO prediction. This is presented in section \ref{sec:nnlo_i}.
Finally, we combine the full NLL calculation with the NNLL contributions computed in the heavy top limit. This is presented in section \ref{sec:nnll}.

\subsection{NLL+NLO with full $M_t$ dependence}
\label{sec:nll}

{\renewcommand{\arraystretch}{1.6}
\begin{table}[t]
\begin{center}
\begin{tabular}{l|c|c|c|c}
$\sqrtS$  & NLO $(\mu_0 = M_{hh}/2)$ & NLL $(\mu_0 = M_{hh}/2)$  & $\f{\delta\text{NLL}}{\text{NLO}}$ $(\mu_0 = M_{hh}/2)$ & $\f{\delta\text{NLL}}{\text{NLO}}$ $(\mu_0 = M_{hh})$
\\[0.5ex]
\hline
\hline
\Tstrut
7~TeV           
& $5.773\,^{+16.2\%}_{-15.1\%}$~fb
& $6.121\,^{+10.9\%}_{-10.3\%}$~fb
& $6.0 \%$ 
& $21.3 \%$ 
\Bstrut\\
\hline
\Tstrut
8~TeV           
& $8.342\,^{+15.7\%}_{-14.6\%}$~fb
& $8.801\,^{+10.9\%}_{-10.2\%}$~fb
& $5.5 \%$ 
& $20.1 \%$ 
\Bstrut\\
\hline
\Tstrut
13~TeV 
& $27.78\,^{+13.8\%}_{-12.8\%}$~fb
& $28.92\,^{+10.7\%}_{-10.1\%}$~fb
& $4.1 \%$ 
& $16.7 \%$ 
\Bstrut\\
\hline
\Tstrut
14~TeV 
& $32.88\,^{+13.5\%}_{-12.5\%}$~fb
& $34.18\,^{+10.7\%}_{-10.1\%} $~fb
& $3.9 \%$  
& $16.3 \%$  
\Bstrut\\
\hline
\Tstrut
27~TeV  
& $127.7\,^{+11.5\%}_{-10.4\%} $~fb
& $131.3\,^{+10.4\%}_{-9.9\%} $~fb
& $2.8 \% $ 
& $13.4 \% $ 
\Bstrut\\
\hline
\Tstrut
100~TeV 
& $1147\,^{+10.7\%}_{-9.9\%} $~fb
& $1166\,^{+11.0\%}_{-9.6\%} $~fb
& $1.7 \% $ 
& $10.2 \% $ 
\Bstrut\\
[0.5ex]\hline
\end{tabular}
\end{center}
\vspace*{-0.5cm}
\caption{\small
Fixed order NLO and resummed NLL+NLO predictions for the Higgs boson pair production total cross section, for different collider energies.
The scale uncertainties are indicated as superscript/subscript. We also present the size of the resummed contribution relative to the NLO result, for both $\mu_0 = M_{hh}/2$ and $\mu_0 = M_{hh}$.
}
\label{table:totalXS}
\end{table}

The results for the total cross section are shown in Table \ref{table:totalXS} for different center-of-mass energies.
We use as the central scale $\mu_0 = M_{hh}/2$, though we also present results for $\mu_0 = M_{hh}$. Scale uncertainties are obtained via the usual 7-point variation.

We can observe that the size of the threshold effects goes down for larger collider energies, as expected from the fact that more energy is available and therefore soft gluon contributions become less dominant.
As it was also observed in the heavy $M_t$ limit, we can appreciate that the size of the threshold corrections is much larger for $\mu_0 = M_{hh}$, ranging from $21.3\%$ at 7~TeV to $10.2\%$ at 100~TeV.
The corresponding values for $\mu_0 = M_{hh}/2$ are $6.0\%$ and $1.7\%$, respectively.
For LHC energies, the soft gluon resummation effects are of the order of $4\%$ for the central scale $\mu_0 = M_{hh}/2$.
We can observe a reduction in the scale uncertainties (except for the 100~TeV predictions, where fixed-order and resummed results are comparable), this reduction being stronger for smaller center-of-mass energies.
In fact, the NLL relative scale uncertainties remain practically unchanged when varying the collider energy, being always about $\pm 10\%$.

In Table \ref{table:ratios} we present the ratio of the central values for the predictions corresponding to $\mu_0 = M_{hh}/2$ and  $\mu_0 = M_{hh}$, both for the fixed-order and resummed results.
We can observe that the variation is substantially smaller in the resummed case, pointing towards a clear improvement in the stability of the cross section when taking into account the all-orders soft gluon effects.

{\renewcommand{\arraystretch}{1.6}
\begin{table}[t]
\begin{center}
\begin{tabular}{l|c|c}
$\sqrtS$  & $\f{\text{NLO}(\mu_0 = M_{hh}/2)}{\text{NLO}(\mu_0 = M_{hh})} -1 $ & $\f{\text{NLL}(\mu_0 = M_{hh}/2)}{\text{NLL}(\mu_0 = M_{hh})} - 1$
\\[0.5ex]
\hline
\hline
\Tstrut
7~TeV           
& $17.9 \%$ 
& $3.0 \%$ 
\Bstrut\\
\hline
\Tstrut
8~TeV           
& $17.1 \%$ 
& $2.9 \%$ 
\Bstrut\\
\hline
\Tstrut
13~TeV 
& $14.7 \%$ 
& $2.3 \%$ 
\Bstrut\\
\hline
\Tstrut
14~TeV 
& $14.3 \%$  
& $2.2 \%$  
\Bstrut\\
\hline
\Tstrut
27~TeV  
& $11.7 \% $ 
& $1.3 \% $ 
\Bstrut\\
\hline
\Tstrut
100~TeV 
& $7.7 \% $ 
& $-0.6 \% $ 
\Bstrut\\
[0.5ex]\hline
\end{tabular}
\end{center}
\vspace*{-0.5cm}
\caption{\small
Ratio between the $\mu_0 = M_{hh}/2$ and $\mu_0 = M_{hh}$ predictions, at NLO and NLL.
}
\label{table:ratios}
\end{table}

We also present NLL predictions (with $\mu_0 = M_{hh}/2$) for the Higgs pair invariant mass $M_{hh}$, at 7~TeV, 13~TeV (Figure \ref{fig:mhh_7_and_13}), 27~TeV and 100~TeV (Figure \ref{fig:mhh_27_and_100}).
The lower plots show the ratio to the NLO result.
We can see that the effect of the resummed contributions becomes larger as the invariant mass of the system increases, which again is expected due to the fact that less energy is available for extra emission.
The increase in the Sudakov factor is however partially compensated by a suppression at large $M_{hh}$ in the NLO virtual corrections entering in $C^{(1)}_{gg}$, leading to a rather mild increase in the tail.
Also here we can clearly observe that the resummation effects decrease with the collider energy.

\begin{figure}
\begin{center}
\includegraphics[width=.49\textwidth]{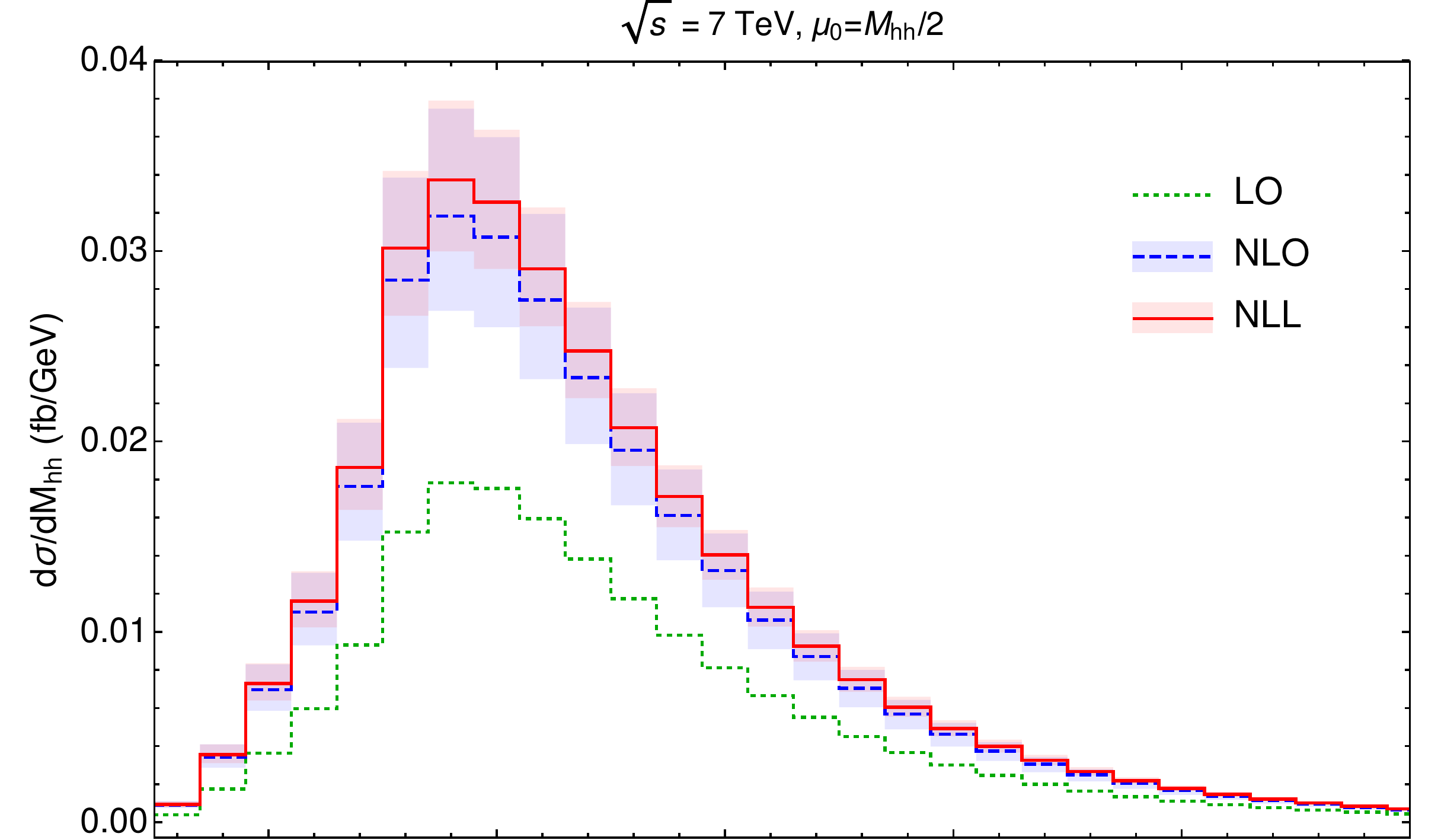}
\hfill
\includegraphics[width=.49\textwidth]{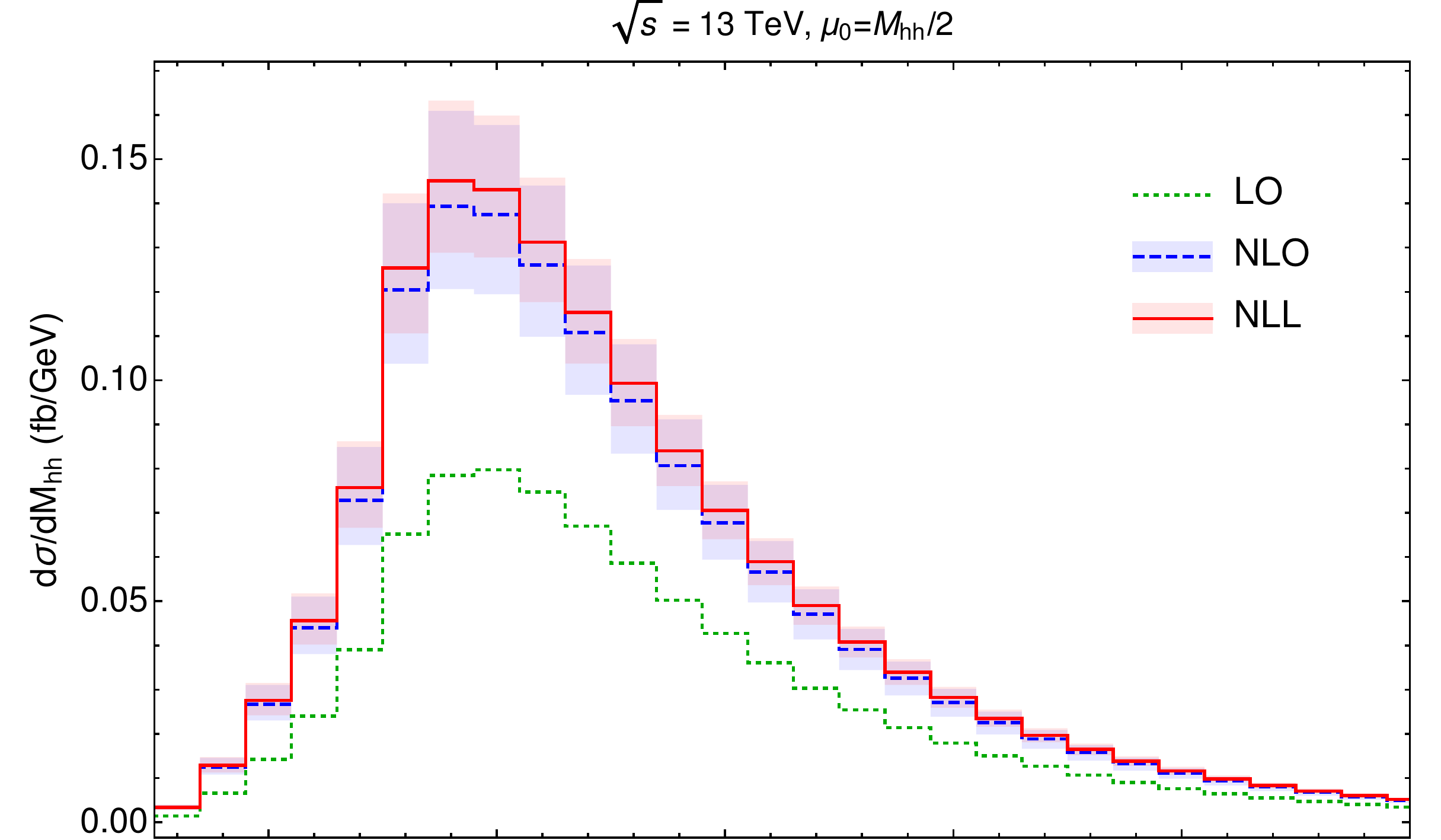}
\\
\includegraphics[width=.49\textwidth]{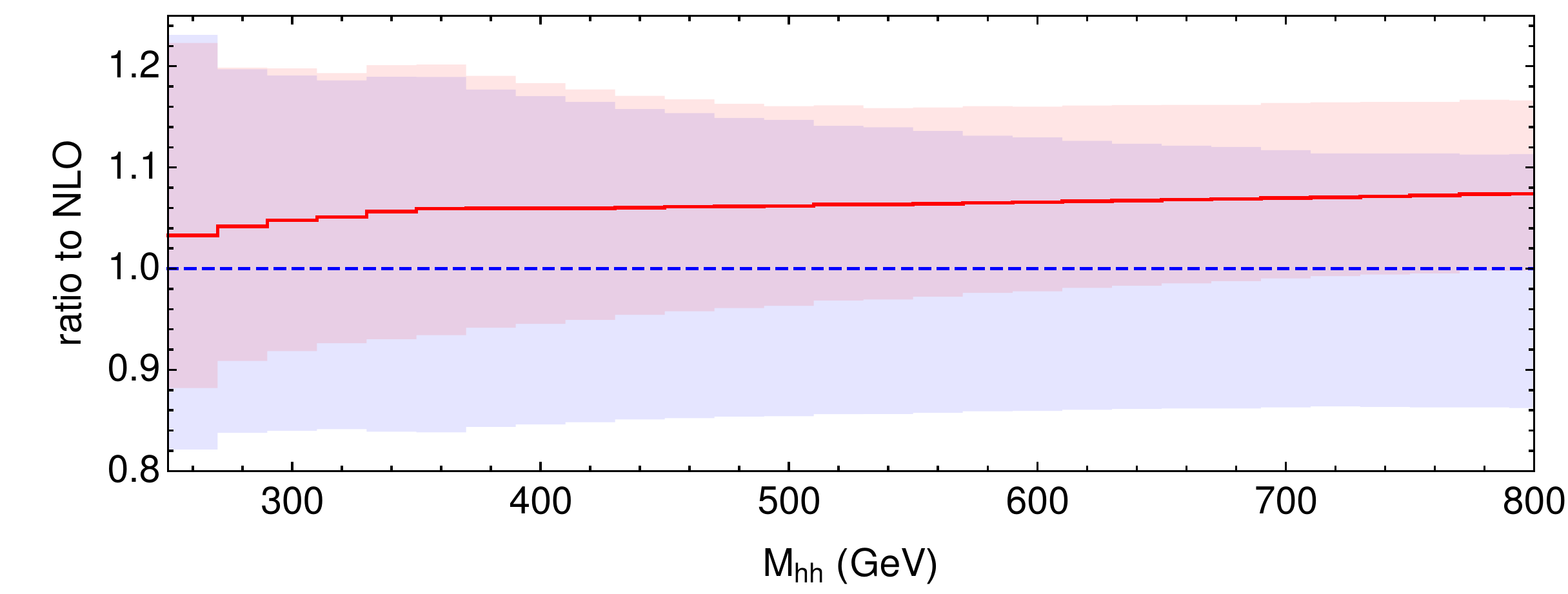}
\hfill
\includegraphics[width=.49\textwidth]{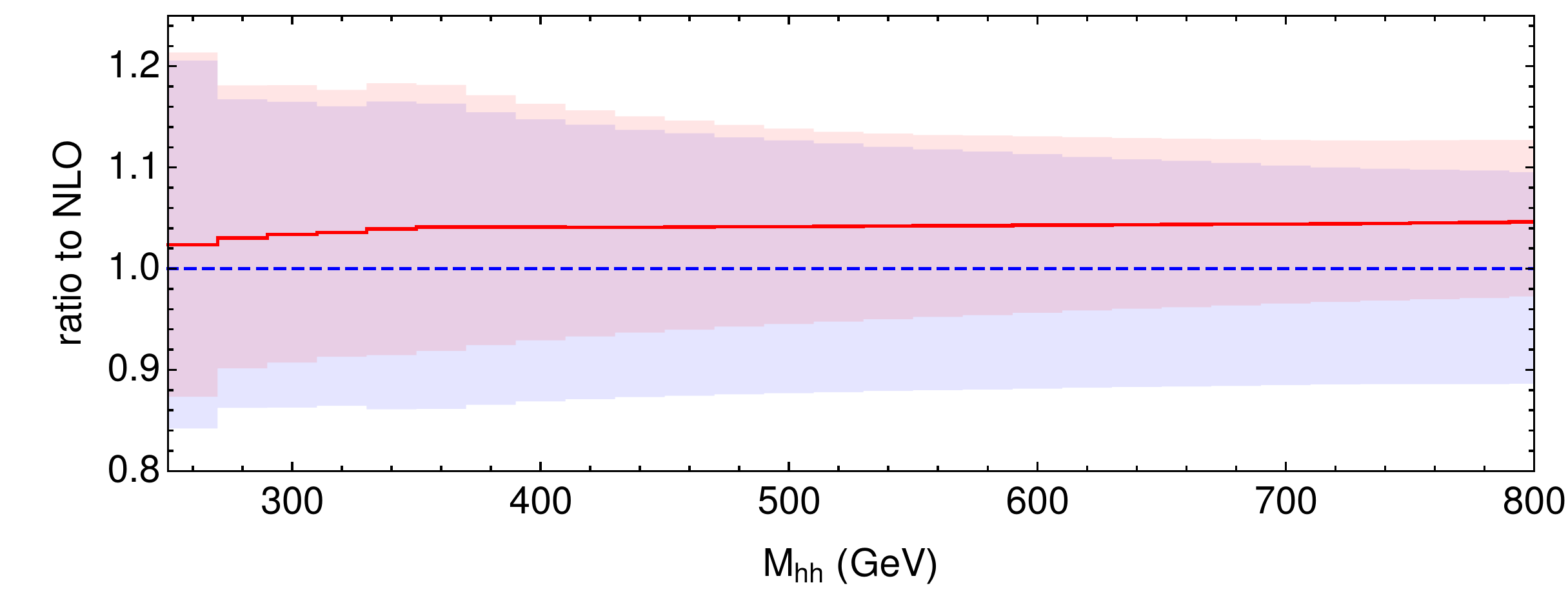}
\end{center}
\vspace{-2ex}
\caption{\label{fig:mhh_7_and_13}\small
Higgs pair invariant mass distribution at LO (green dotted), NLO (blue dashed) and NLL+NLO (red solid), for collider energies of 7~TeV (left) and 13~TeV (right).
The lower panel shows the ratio to the NLO result.
The bands indicate the NLO and NLL+NLO scale uncertainties.
}
\end{figure}

\begin{figure}
\begin{center}
\includegraphics[width=.49\textwidth]{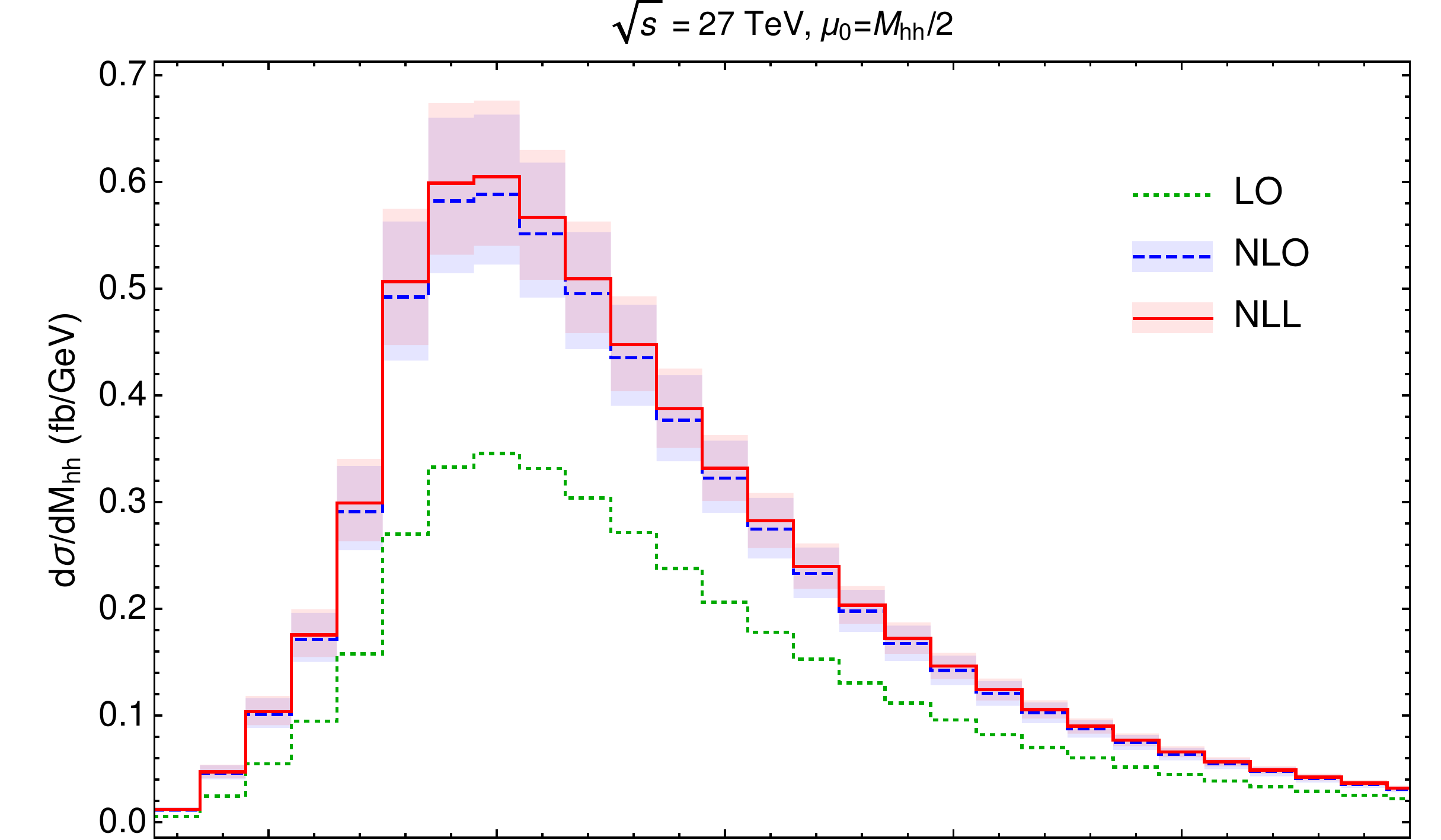}
\hfill
\includegraphics[width=.49\textwidth]{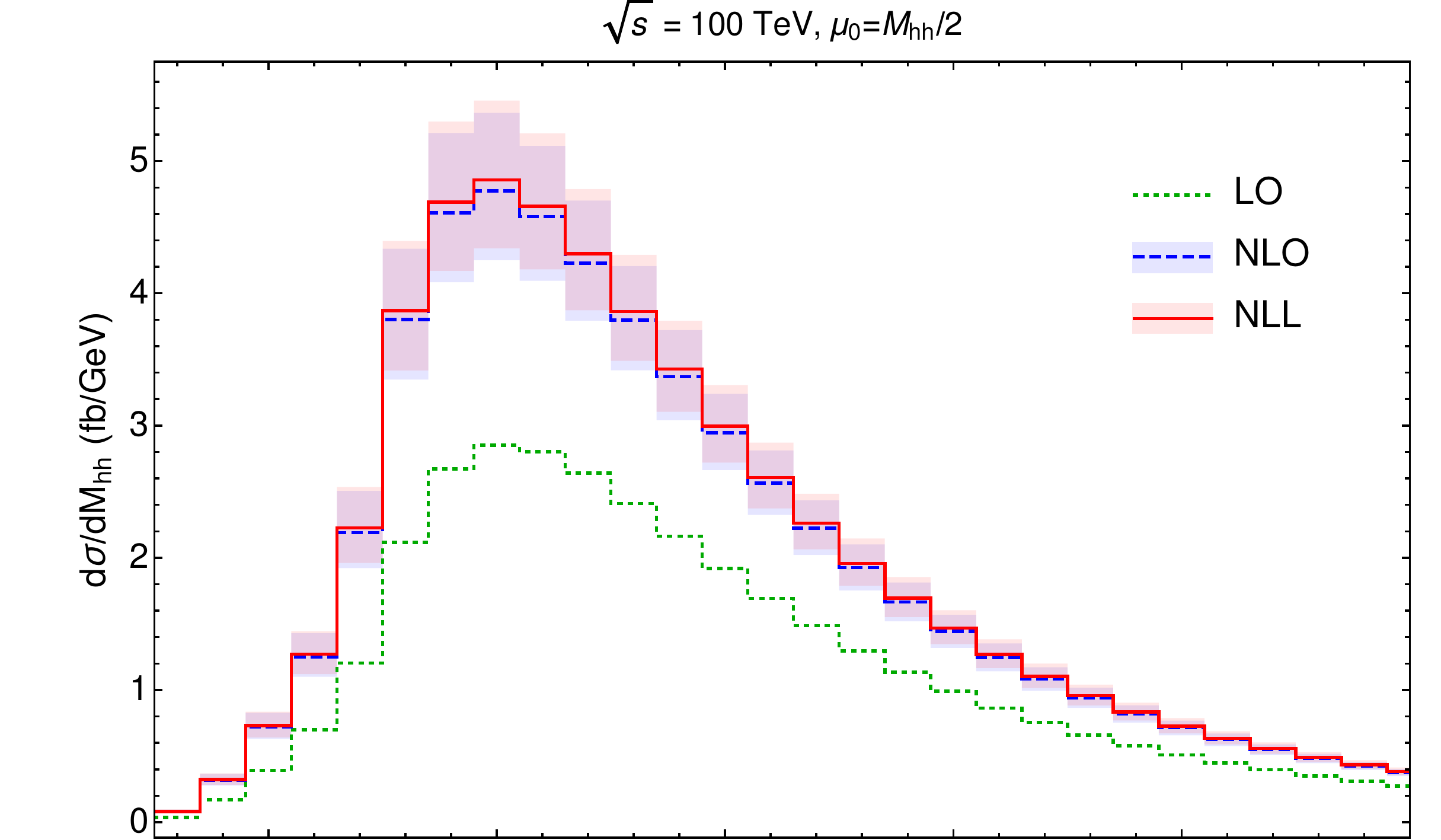}
\\
\includegraphics[width=.49\textwidth]{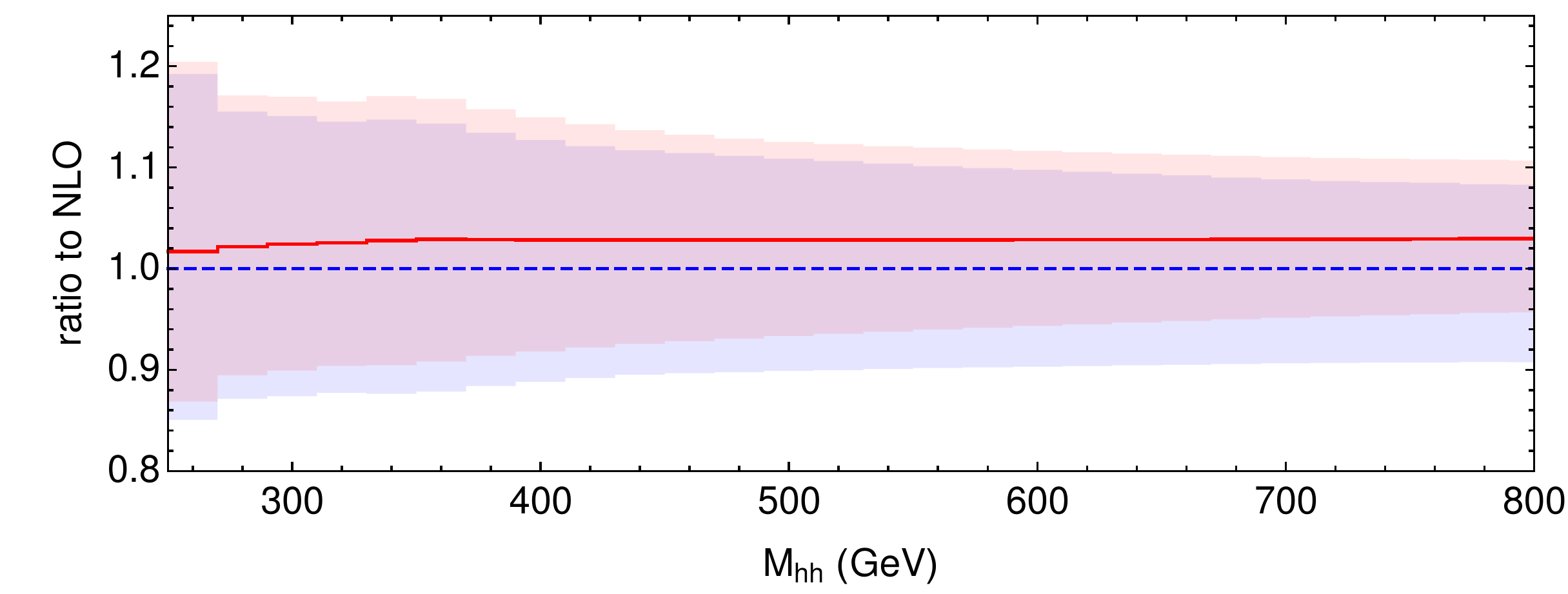}
\hfill
\includegraphics[width=.49\textwidth]{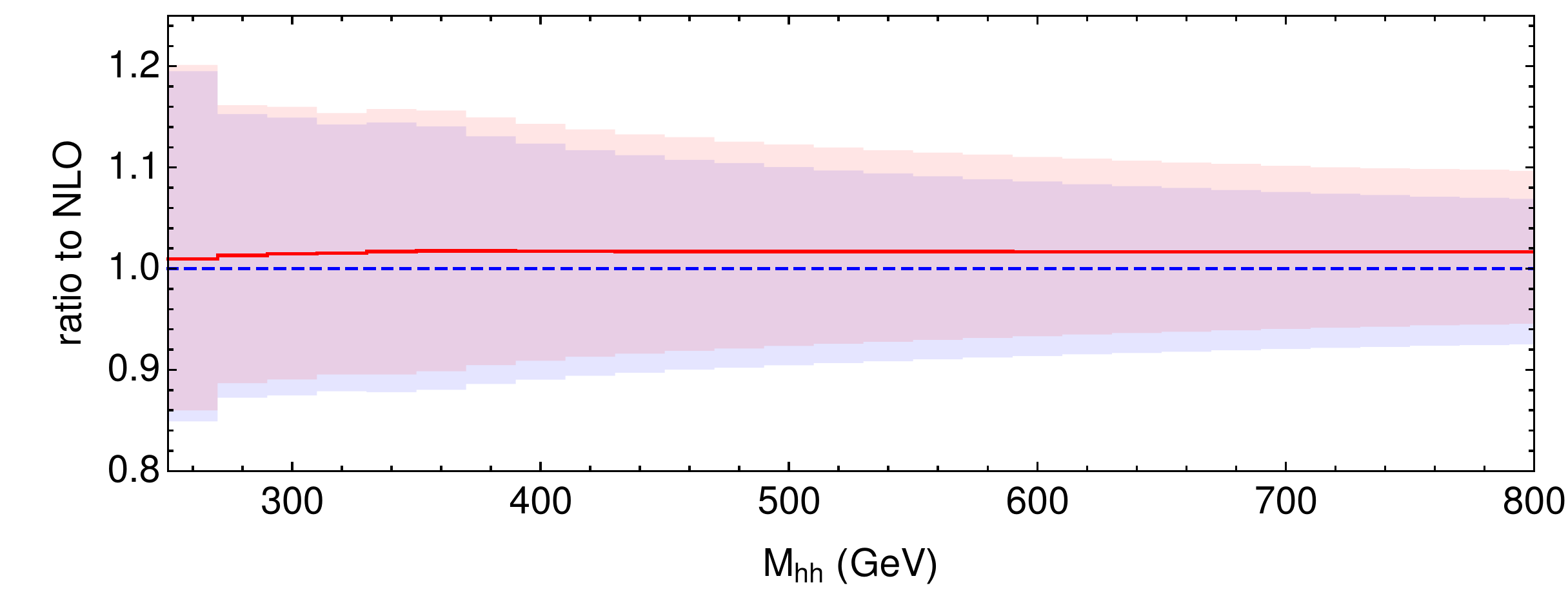}
\end{center}
\vspace{-2ex}
\caption{\label{fig:mhh_27_and_100}\small
Higgs pair invariant mass distribution at LO (green dotted), NLO (blue dashed) and NLL+NLO (red solid), for collider energies of 27~TeV (left) and 100~TeV (right).
The lower panel shows the ratio to the NLO result.
The bands indicate the NLO and NLL+NLO scale uncertainties.
}
\end{figure}

It is interesting to compare our results with the ones obtained in the heavy-$M_t$ limit \cite{deFlorian:2015moa}.
In order to do so, we present in Figure \ref{fig:ratio_energies} the ratio between the NLL and NLO predictions as a function of $M_{hh}$ for different collider energies, both in the full theory and in the HTL.
We can observe that there are clear differences in the shape, with the results with full $M_t$ dependence growing faster for lower invariant masses but showing a relative suppression with respect to the large-$M_t$ results in the tail.
Still, this difference in the $M_{hh}$ spectrum between the two predictions is of the order of $\pm 1\%$, and it is moderate compared to the overall effect of the resummed contributions.
This indicates certain stability in the $M_t$ dependence of the threshold effects, and therefore the lack of full $M_t$ dependence at NNLL should lead to a rather small residual uncertainty due to missing finite-$M_t$ effects

\begin{figure}
\begin{center}
\includegraphics[width=.65\textwidth]{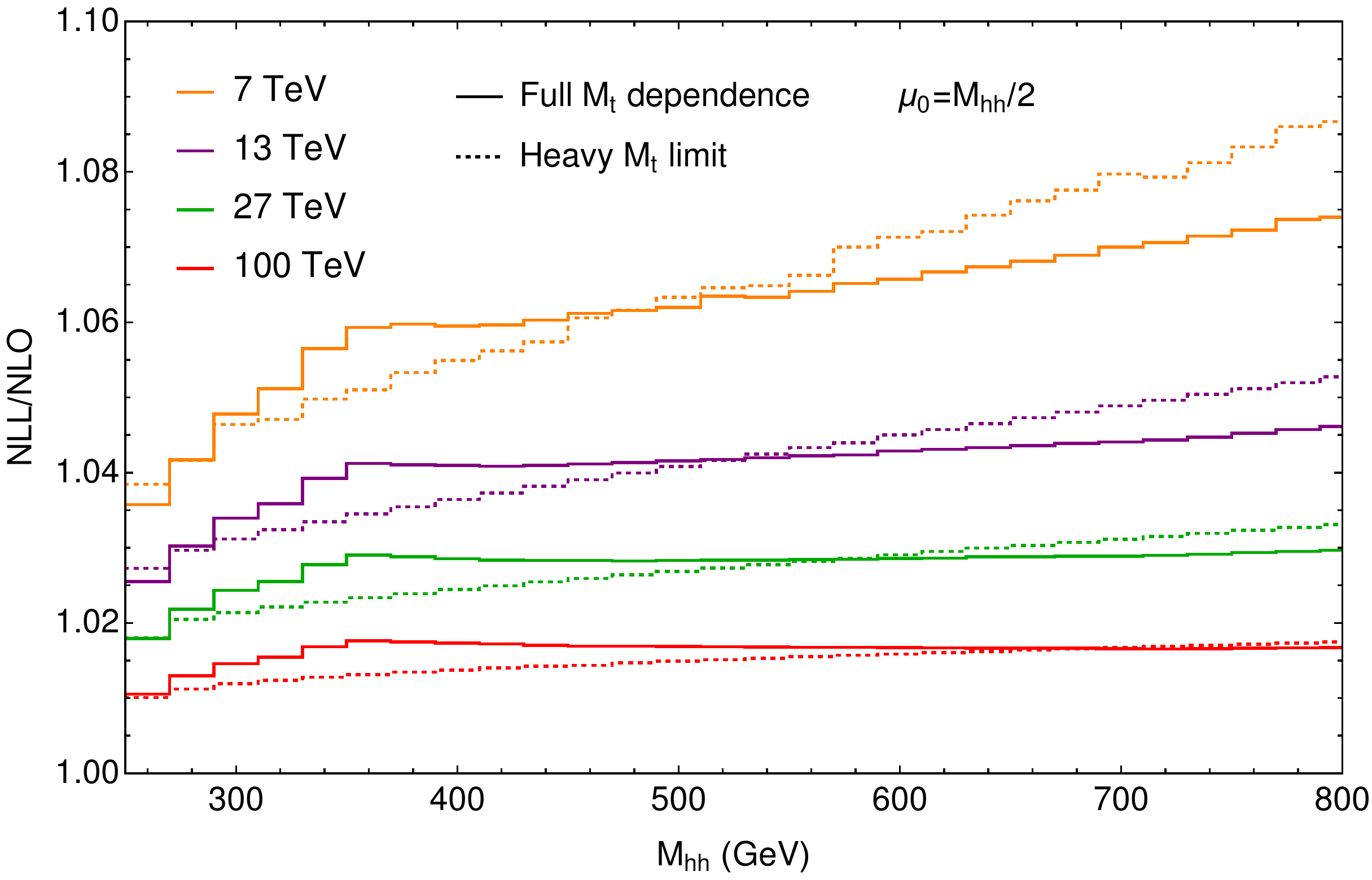}
\hfill
\end{center}
\vspace{-2ex}
\caption{\label{fig:ratio_energies}\small
Ratio between the NLL+NLO and NLO predictions, as a function of the Higgs pair invariant mass and for different collider energies.
The solid curves show the results with full $M_t$ dependence, while the dashed ones correspond to the large $M_t$ limit.
}
\end{figure}

\subsection{Improved \nnloFT}
\label{sec:nnlo_i}

As it was mentioned in the previous section, the NLL+NLO results represent the most advanced prediction available for double Higgs production in the full theory.
However, higher order corrections are still sizeable and therefore they need to be included in order to obtain accurate results, even if they are known only in an approximated way.
The best fixed order prediction available in the literature is the so-called \nnloFT \cite{Grazzini:2018bsd}, which is obtained by working in the heavy $M_t$ limit but improved via a reweighting technique in order to account for finite $M_t$ effects. In particular, the \nnloFT includes the full double-real loop induced squared matrix elements.

Before presenting combined NNLL+\nnloFT predictions in the following section, it is worth to discuss possible improvements to the approximated NNLO result of Ref.~\cite{Grazzini:2018bsd} based on the knowledge of the full NLL+NLO result.
Expanding the NLL+NLO results to ${\cal O}(\as^2)$ --where an overall $\as^2$ from the Born cross section is understood--, we can obtain the exact threshold enhanced contributions proportional to $\as^2 \ln^2 N$\footnote{Contributions proportional to $\as^2 \ln^3 N$ and $\as^2 \ln^4 N$ are already obtained in an exact way at LL, and are also reproduced with full $M_t$ dependence by the \nnloFT.}.
Even if it features the full double-real corrections, these contributions are obtained only within the (Born-improved) heavy $M_t$ limit in the \nnloFT, because of the approximation performed in the real-virtual piece of the calculation.
Therefore, we can define an improved \nnloFT (denoted as \nnloFTi) in the following way\footnote{Besides having the full $M_t$ dependence in the $\as^2 \ln^2 N$ term, the \nnloFTi differs from the \nnloFT result also in the term proportional to $\as^2 \ln N$, though in this case the full $M_t$ dependence is only in those contributions generated by the NLL resummation.}
\beq
\sigma^\text{NNLO}_\text{FTa-i} =
\sigma^\text{NNLO}_\text{FTa}
+ \left( \sigma^\text{NLL}
-\sigma^\text{NLL}_\text{HTL}
\right)
\Big |_{\text{only }{\cal O}(\as^2)}\,.
\eeq

In Table \ref{table:nnlo} we show the comparison between the \nnloFT and \nnloFTi predictions for the total cross section.
We can observe that the difference is very small, being always below $0.5\%$.
Even if this does not represent a proof of the accuracy of the \nnloFT, the smallness of this effect points in this direction, and the difference is largely included within the estimated $M_t$ uncertainty reported in Ref.~\cite{Grazzini:2018bsd}.

{\renewcommand{\arraystretch}{1.6}
\begin{table}[t]
\begin{center}
\begin{tabular}{l|c|c|c|c|c|c}
\hspace*{-0.2cm}
$\sqrtS$  & 7~TeV & 8~TeV & 13~TeV & 14~TeV & 27~TeV & 100~TeV
\\[0.5ex]
\hline
\hline
\Tstrut
\hspace*{-0.2cm}
\nnloFT [fb]
& $6.572\,^{+3.0\%}_{-6.5\%} $
& $9.441\,^{+2.8\%}_{-6.1\%} $
& $31.05\,^{+2.2\%}_{-5.0\%} $
& $36.69\,^{+2.1\%}_{-4.9\%} $ 
& $139.9\,^{+1.3\%}_{-3.9\%} $ 
& $1224\,^{+0.9\%}_{-3.2\%} $ 
\Bstrut\\
\hline
\Tstrut
\hspace*{-0.2cm}
\nnloFTi [fb] \hspace*{-0.2cm}
& $6.547\,^{+3.4\%}_{-6.9\%} $
& $9.406\,^{+3.2\%}_{-6.5\%} $
& $30.95\,^{+2.9\%}_{-5.5\%} $
& $36.57\,^{+2.7\%}_{-5.3\%} $ 
& $139.5\,^{+2.4\%}_{-4.3\%} $ 
& $1221\,^{+2.0\%}_{-3.2\%} $ 
\Bstrut\\
\hline
\Tstrut
\hspace*{-0.2cm}
\nnllFTi [fb] \hspace*{-0.2cm}
& $6.633\,^{+3.8\%}_{-3.8\%} $
& $9.515\,^{+3.7\%}_{-3.7\%} $
& $31.18\,^{+3.3\%}_{-3.6\%} $
& $36.83\,^{+3.3\%}_{-3.5\%} $ 
& $140.1\,^{+3.0\%}_{-3.3\%} $ 
& $1223\,^{+2.4\%}_{-2.8\%} $ 
\Bstrut\\
\hline
\Tstrut
\hspace*{-0.2cm}
$\f{\delta\text{NNLL}_\text{FTa-i}}{\text{NNLO}_\text{FTa-i}}$
\hspace*{-0.2cm}
& $ 1.3 \% $
& $ 1.2 \% $
& $ 0.8 \% $
& $ 0.7 \% $ 
& $ 0.4 \% $ 
& $ 0.1 \% $ 
\Bstrut\\
[0.5ex]\hline
\end{tabular}
\end{center}
\vspace*{-0.5cm}
\caption{\small
Total Higgs boson pair production cross sections at hadron colliders at \nnloFT, \nnloFTi and NNLL$+$\nnloFTi (labeled \nnllFTi for brevity), for different center of mass energies. All the results correspond to the central scale $\mu_0 = M_{hh}/2$.
}
\label{table:nnlo}
\end{table}

In Figure \ref{fig:mhh_13_nnlo} we present the Higgs boson pair invariant mass distribution for both NNLO approximations, for a collider energy of 13~TeV. We can observe that the difference between them is again very small in the whole invariant mass range, slowly growing with $M_{hh}$ but always within the scale uncertainties.
This behavior is not surprising since the \nnloFT is expected to be less accurate for large values of $M_{hh}$, and also because the difference between \nnloFT and \nnloFTi is only in threshold enhanced terms, which become more relevant for larger invariant masses.
We can also observe that the scale uncertainties are larger for the \nnloFTi in the tail, being the central value corresponding to $\mu_0 = M_{hh}/2$ in the middle of the uncertainty band, while for the \nnloFT it is located close to the upper limit.
This fact reflects in the slightly larger scale uncertainties for the \nnloFTi total cross section that can also be observed in Table \ref{table:nnlo}.

\begin{figure}
\begin{center}
\includegraphics[width=.55\textwidth]{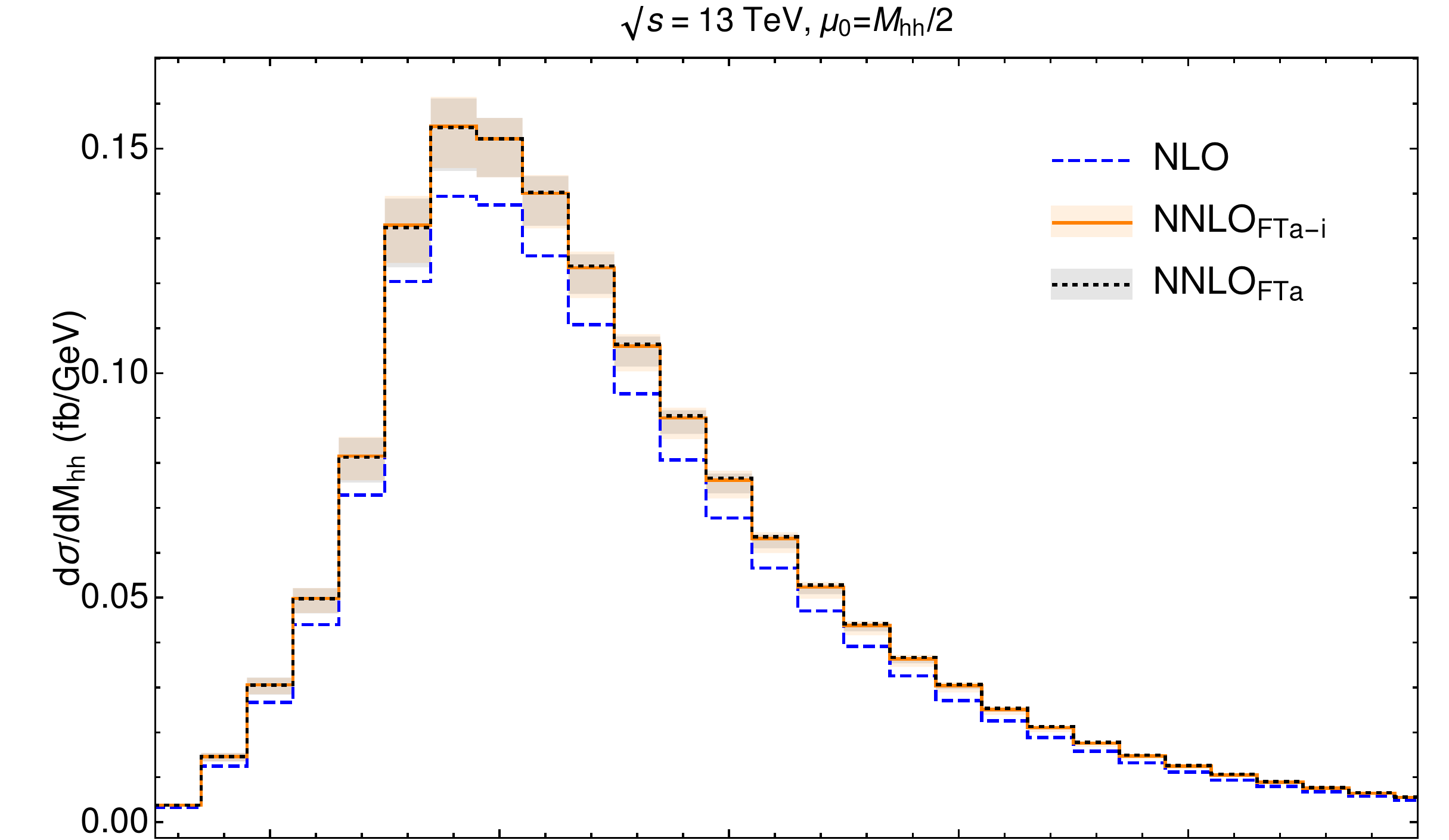}
\hfill
\\
\includegraphics[width=.55\textwidth]{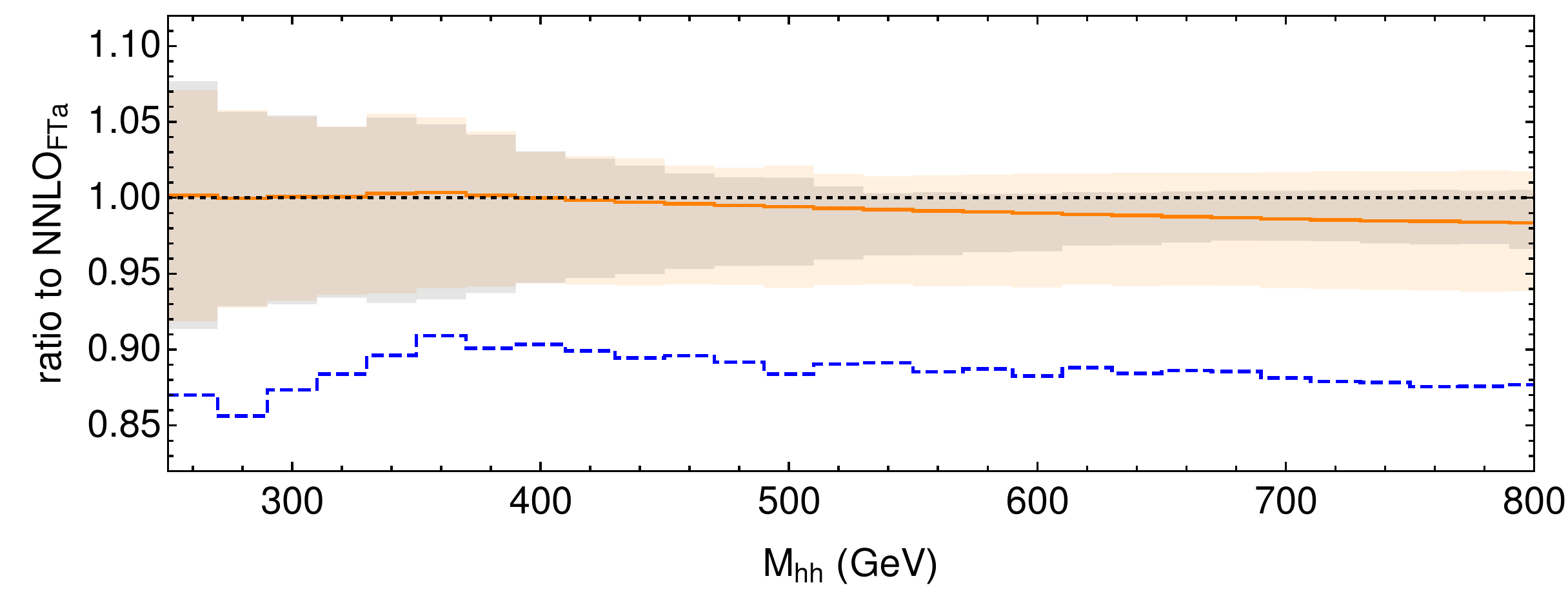}
\hfill
\end{center}
\vspace{-2ex}
\caption{\label{fig:mhh_13_nnlo}\small
Higgs pair invariant mass distribution at NLO (blue dashed), \nnloFT (black solid) and \nnloFTi (orange dotted), for a collider energy of 13~TeV.
The lower panel shows the ratio to the \nnloFT result.
The bands indicate the \nnloFT and \nnloFTi scale uncertainties.
}
\end{figure}

In summary, both for the total cross section and the invariant mass distribution we find that the differences between the \nnloFT and \nnloFTi predictions are well within the estimated uncertainties inherent to these approximations.

\subsection{NNLL resummation}
\label{sec:nnll}

We present now the NNLL predictions. In order to account for the NLL contributions with full $M_t$ dependence, we add the difference between the full theory and HTL predictions at NLL.
Specifically, defining
\beq
\sigma^\text{NNLL'}
=
\sigma^\text{NNLL}_\text{HTL} + \sigma^\text{NLL} - \sigma^\text{NLL}_\text{HTL}\,,
\eeq
we have that our NNLL+\nnloFTi cross section is given by
\beq
\sigma^\text{NNLL+NNLO}_\text{FTa-i}
=
\sigma^\text{NNLL'} - \sigma^\text{NNLL'}\Big |_{{\cal O}(\as^2)} + \sigma^\text{NNLO}_\text{FTa-i}\,.
\eeq
For the sake of brevity, we will denote this result \nnllFTi.
Note that the NNLL result is matched to the \nnloFTi prediction instead of \nnloFT, though as it was seen in the previous section the difference between the two is very small.

In Table \ref{table:nnlo} we present the \nnllFTi predictions for the total cross section, for $\mu_0 = M_{hh}/2$.
We can observe that the resummed contributions result in a small increase with respect to the \nnloFTi result, ranging from $1.3\%$ at 7~TeV to $0.1\%$ at 100~TeV, and being around $0.8\%$ at the LHC.
Again, the effect is much larger for the central scale $\mu_0 = M_{hh}$, where for instance the increase in the total cross section at 13~TeV is above $8\%$.

From Table \ref{table:nnlo} we can also compare the NNLL predictions with the \nnloFT results of Ref.~\cite{Grazzini:2018bsd}. We can observe that the increase due to the resummed contributions is partially compensated with the existing decrease from the \nnloFT to the \nnloFTi predictions, accidentally making the difference between the \nnloFT and \nnllFTi results even smaller.
The largest difference between these two predictions is in the scale uncertainties, which are comparable in size but turn out to be more symmetric for the \nnllFTi result.

{\renewcommand{\arraystretch}{1.6}
\begin{table}[t]
\begin{center}
\begin{tabular}{l|c|c}
$\sqrtS$  & $\f{\text{NNLO}_\text{FTa-i}(\mu_0 = M_{hh}/2)}{\text{NNLO}_\text{FTa-i}(\mu_0 = M_{hh})} -1 $ & $\f{\text{NNLL}_\text{FTa-i}(\mu_0 = M_{hh}/2)}{\text{NNLL}_\text{FTa-i}(\mu_0 = M_{hh})} - 1$
\\[0.5ex]
\hline
\hline
\Tstrut
7~TeV           
& $7.4 \%$ 
& $-1.3 \%$ 
\Bstrut\\
\hline
\Tstrut
8~TeV           
& $7.0 \%$ 
& $-1.3 \%$ 
\Bstrut\\
\hline
\Tstrut
13~TeV 
& $5.9 \%$ 
& $-1.3 \%$ 
\Bstrut\\
\hline
\Tstrut
14~TeV 
& $5.6 \%$  
& $-1.4 \%$  
\Bstrut\\
\hline
\Tstrut
27~TeV  
& $4.5 \% $ 
& $-1.6 \% $ 
\Bstrut\\
\hline
\Tstrut
100~TeV 
& $2.8 \% $ 
& $-2.1 \% $ 
\Bstrut\\
[0.5ex]\hline
\end{tabular}
\end{center}
\vspace*{-0.5cm}
\caption{\small
Ratio between the $\mu_0 = M_{hh}/2$ and $\mu_0 = M_{hh}$ predictions, at \nnloFTi and \nnllFTi.
}
\label{table:ratios_nnll}
\end{table}

In Table \ref{table:ratios_nnll} we compare the fixed order \nnloFTi and resummed \nnllFTi predictions for the scale choices $\mu_0 = M_{hh}/2$ and $\mu_0 = M_{hh}$.
In accordance with what was observed at NLO and NLL, we can see that the fixed order results present a larger variation in the central value when changing the renormalization and factorization scales, while the resummed results show a better stability.
Again, this effect is less strong when we increase the collider energy.

Finally, in Figures \ref{fig:mhh_7_and_13_nnll} and \ref{fig:mhh_27_and_100_nnll} we present the Higgs pair invariant mass distribution at different collider energies.
We can see again that the threshold effects increase with $M_{hh}$ by comparing the \nnloFTi and \nnllFTi curves.
We observe that also at a differential level that the difference between the \nnloFT and \nnllFTi predictions is very small, being below or around $1\%$ in the mass range under study.
The difference in the scale uncertainty bands between these two predictions can also be appreciated, specially in the tail.

\begin{figure}
\begin{center}
\includegraphics[width=.49\textwidth]{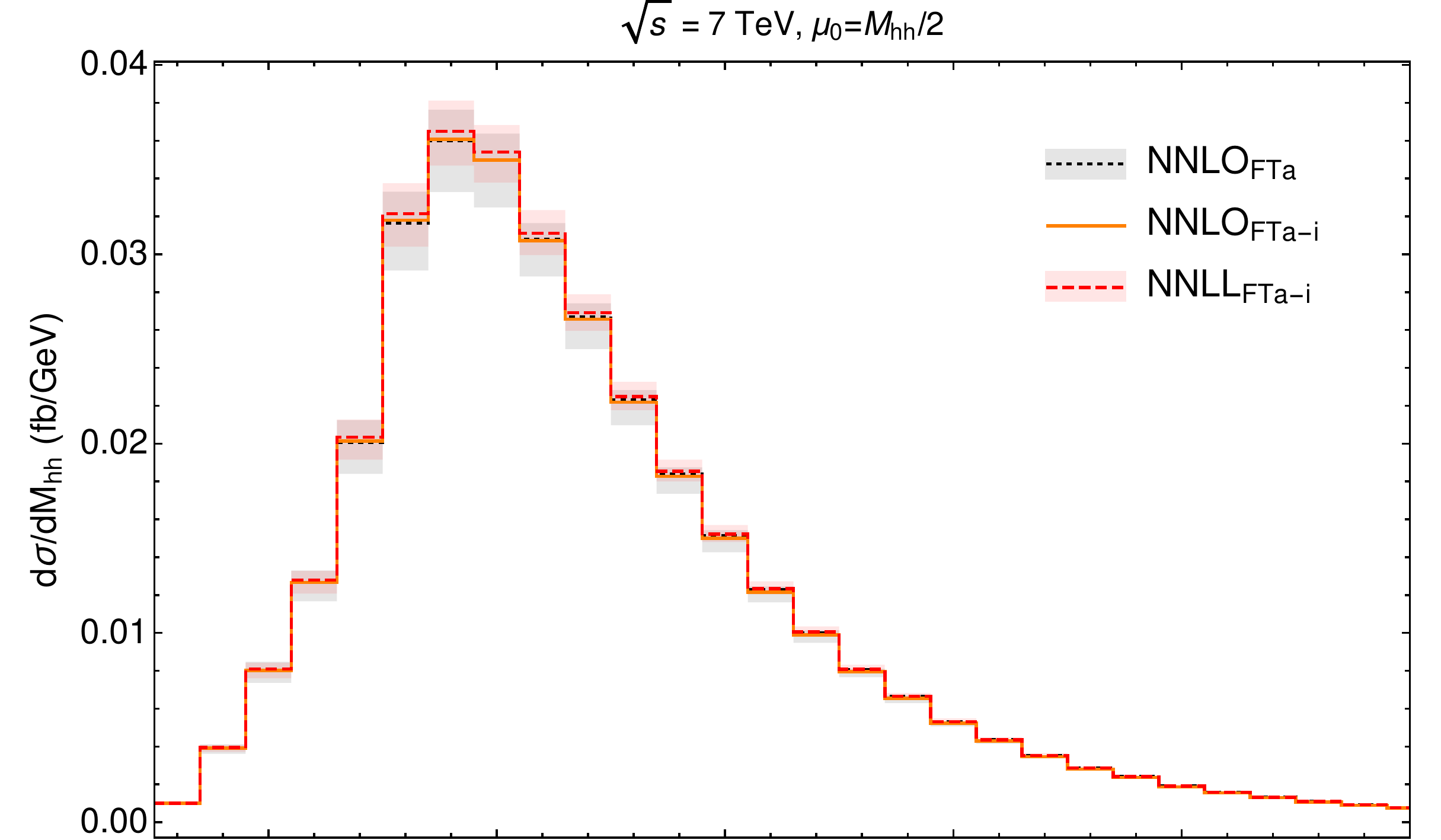}
\hfill
\includegraphics[width=.49\textwidth]{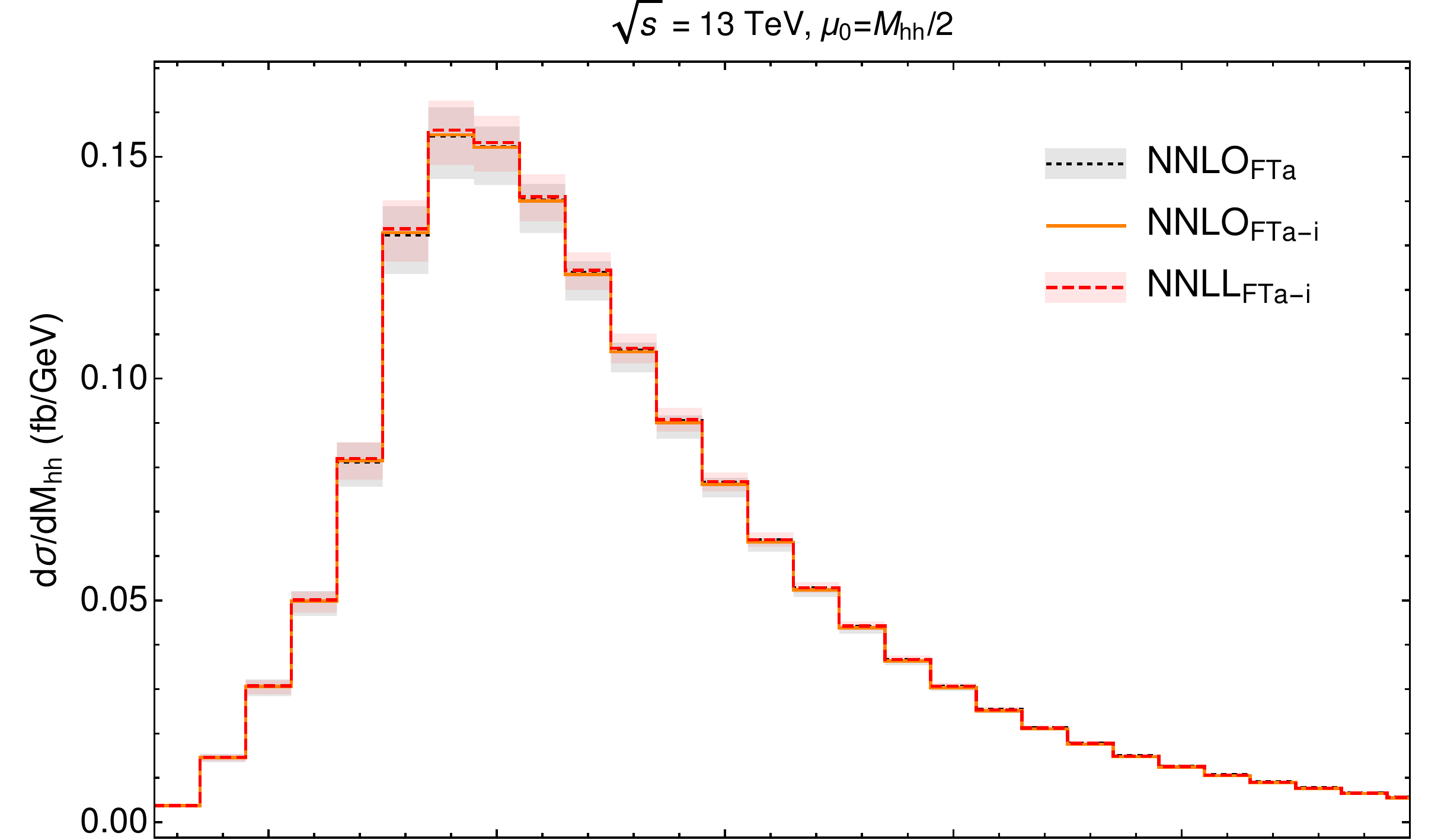}
\\
\includegraphics[width=.49\textwidth]{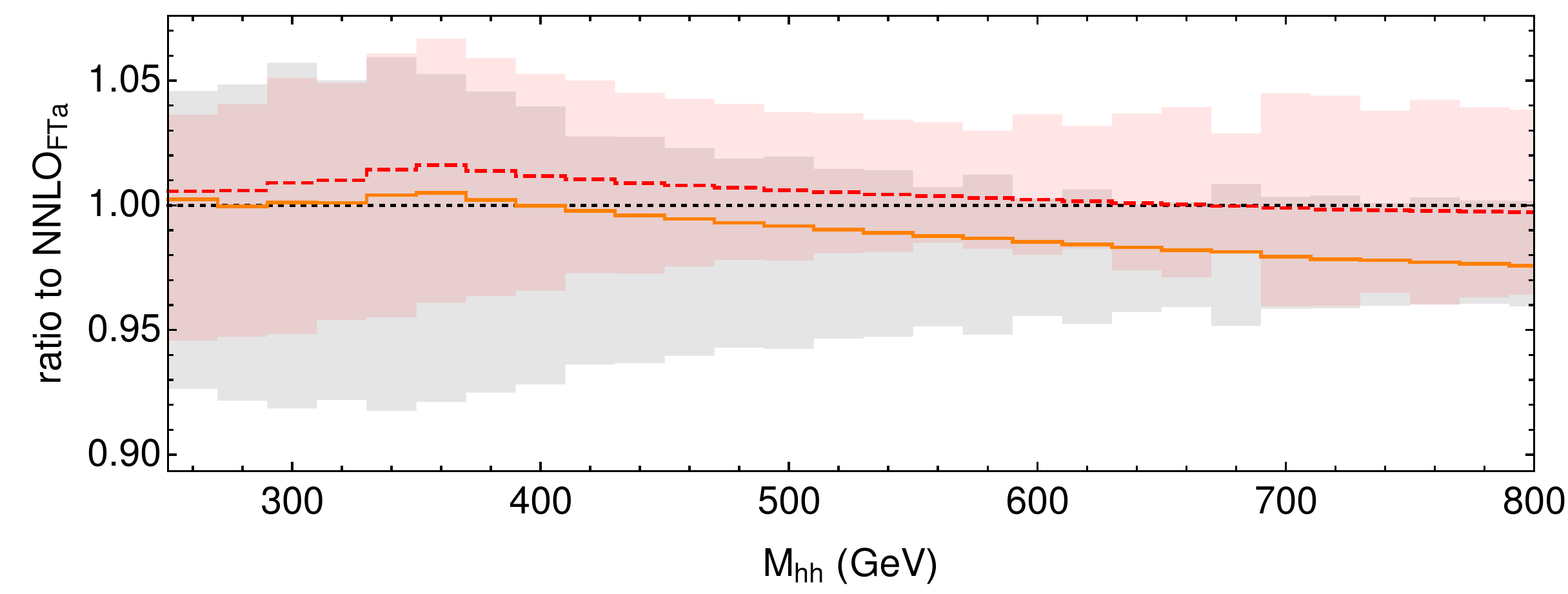}
\hfill
\includegraphics[width=.49\textwidth]{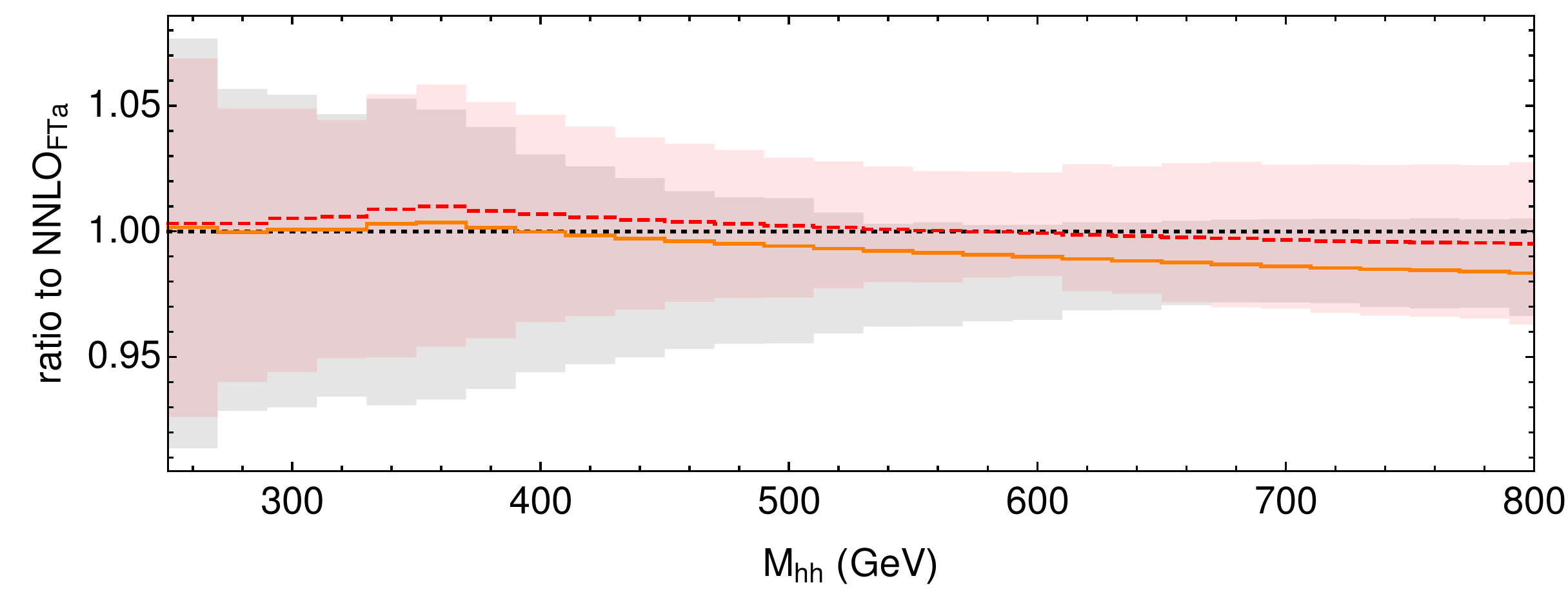}
\end{center}
\vspace{-2ex}
\caption{\label{fig:mhh_7_and_13_nnll}\small
Higgs pair invariant mass distribution at \nnloFT (black dotted), \nnloFTi (orange solid) and \nnllFTi (red dashed), for a collider energy of 7~TeV (left) and 13~TeV (right).
The lower panel shows the ratio to the \nnloFT result.
The bands indicate the \nnloFT and \nnllFTi scale uncertainties.
}
\end{figure}

\begin{figure}
\begin{center}
\includegraphics[width=.49\textwidth]{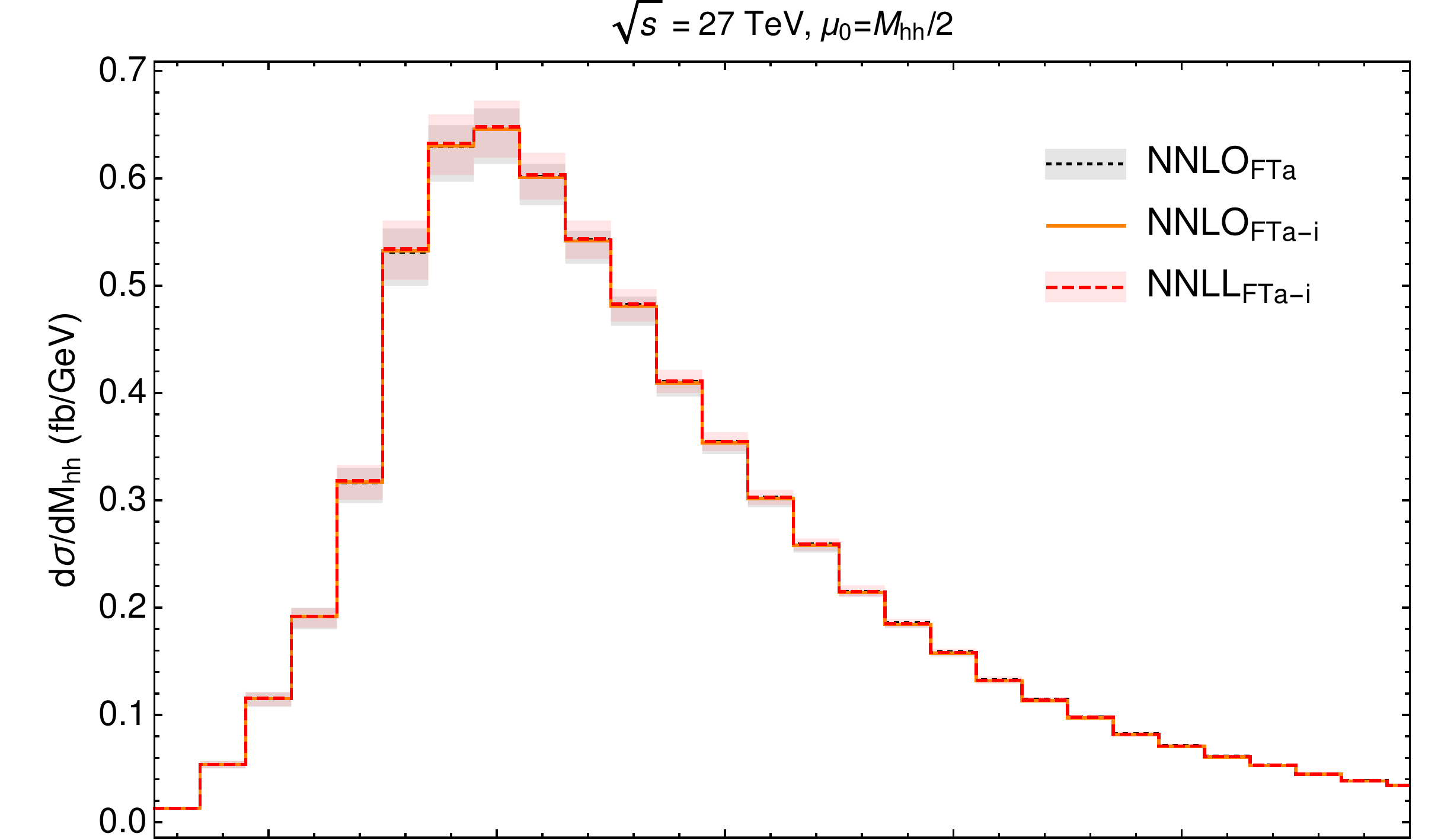}
\hfill
\includegraphics[width=.49\textwidth]{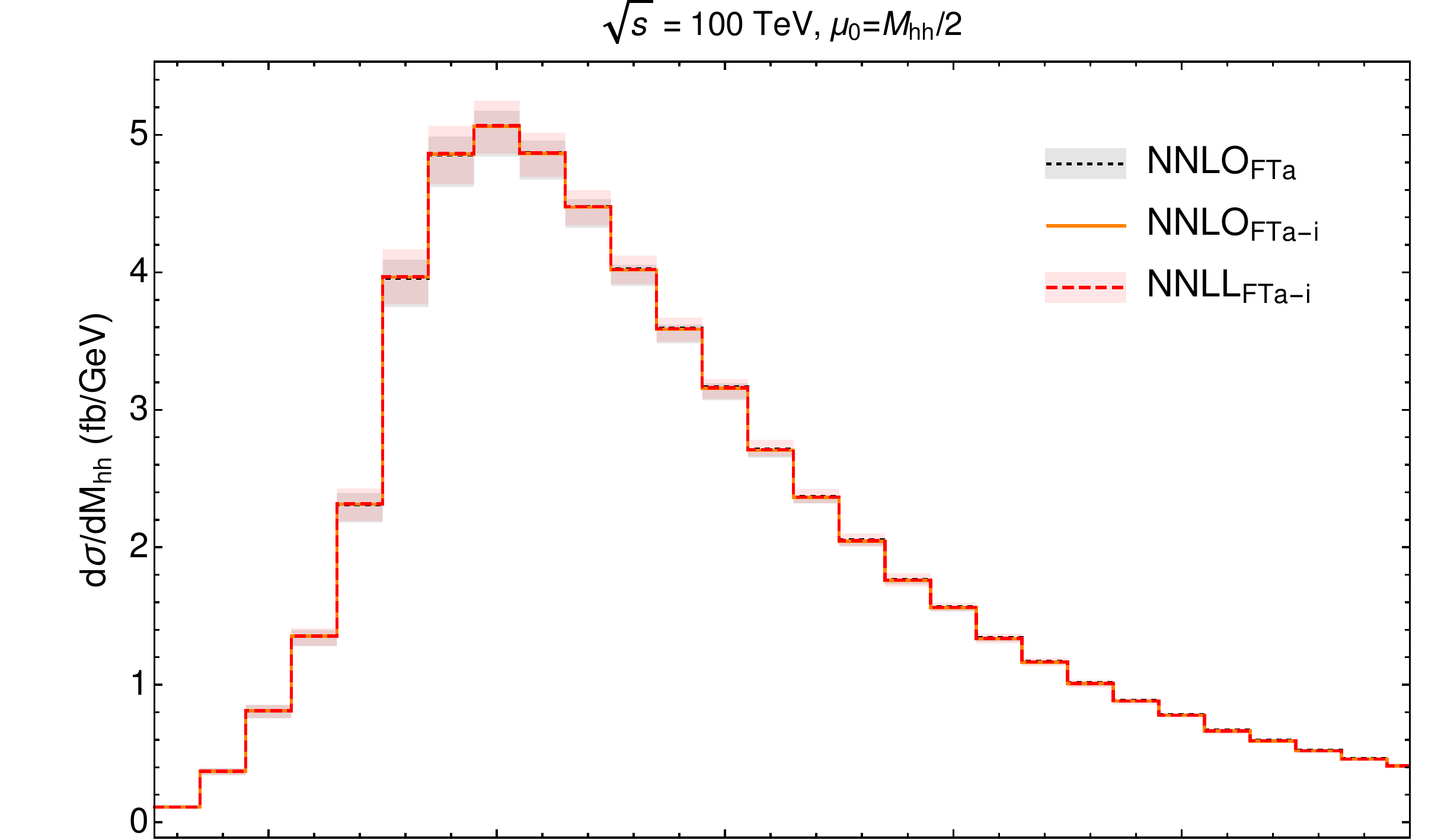}
\\
\includegraphics[width=.49\textwidth]{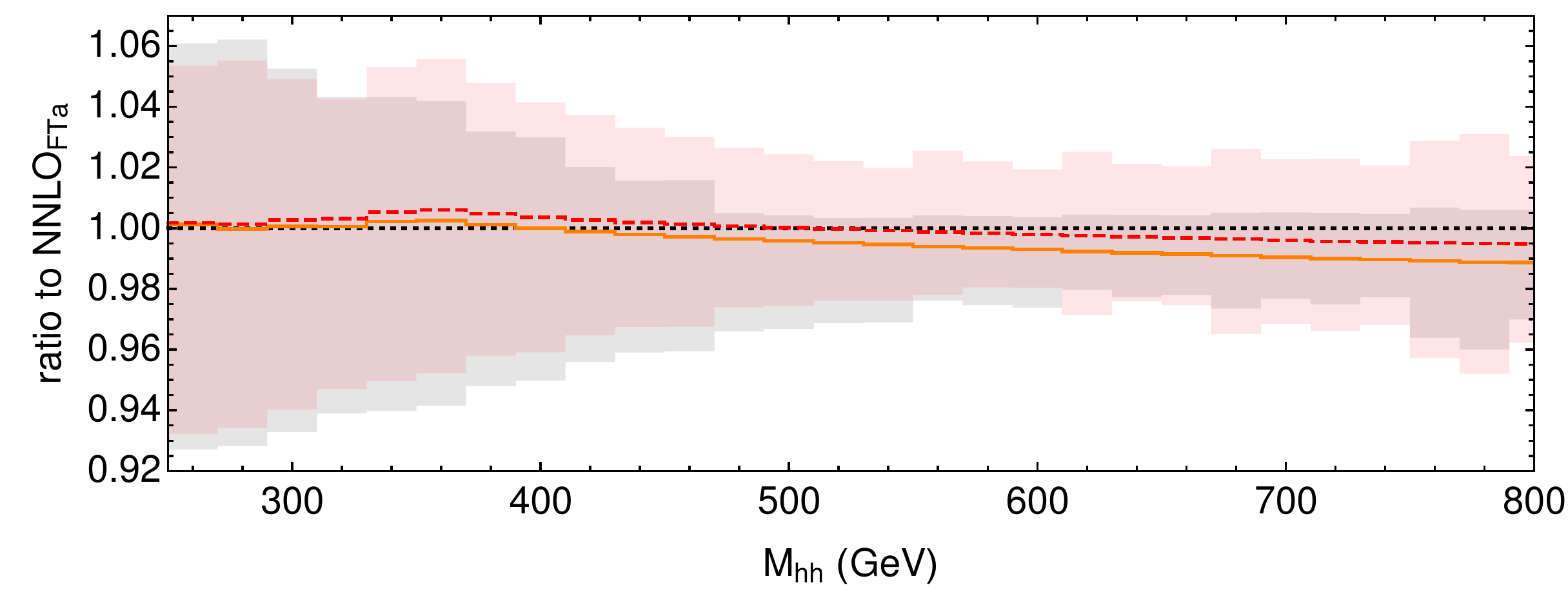}
\hfill
\includegraphics[width=.49\textwidth]{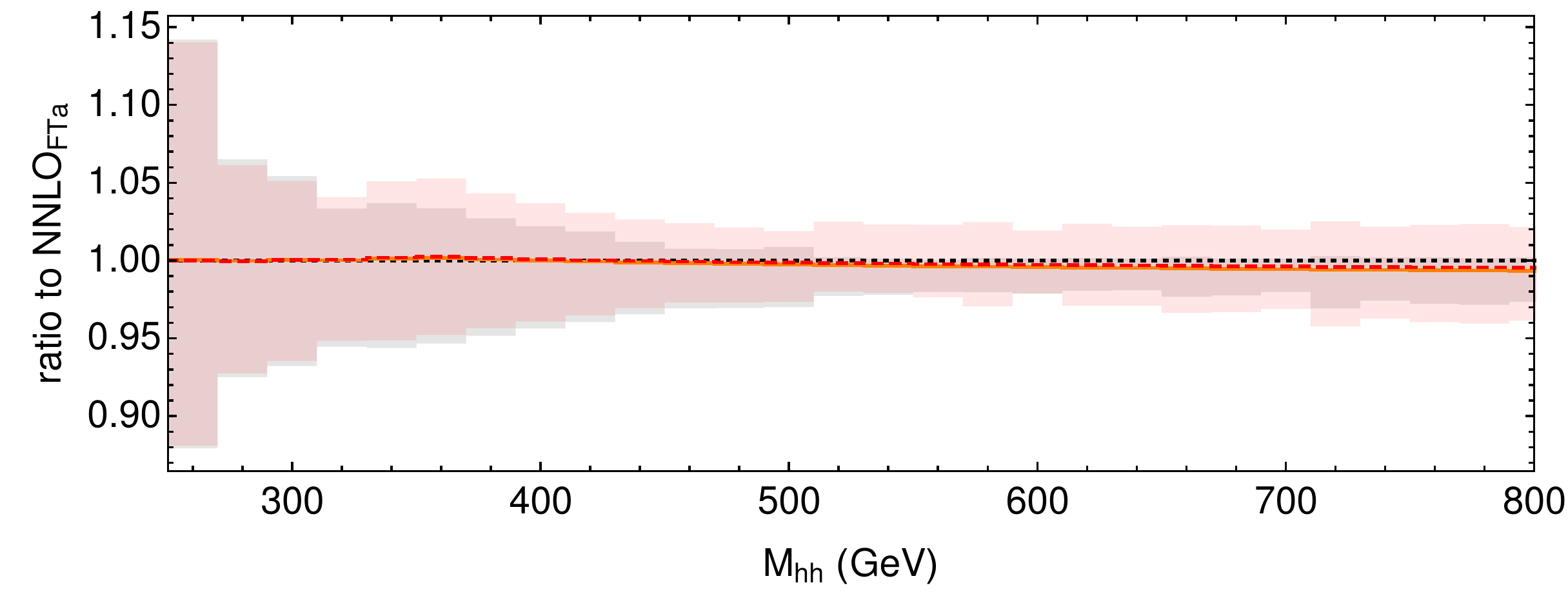}
\end{center}
\vspace{-2ex}
\caption{\label{fig:mhh_27_and_100_nnll}\small
Higgs pair invariant mass distribution at \nnloFT (black dotted), \nnloFTi (orange solid) and \nnllFTi (red dashed), for a collider energy of 27~TeV (left) and 100~TeV (right).
The lower panel shows the ratio to the \nnloFT result.
The bands indicate the \nnloFT and \nnllFTi scale uncertainties.
}
\end{figure}

In conclusion, the difference between the resummed \nnllFTi prediction and the \nnloFT result turns out to be small for $\mu_0 = M_{hh}/2$ compared to the size of the theoretical uncertainties, except only for the effect in the shape of the scale uncertainty bands.
The small impact of the all orders soft gluon resummation is an indication of the good control over the perturbative expansion.

\section{Summary}
\label{sec:conclusions}

In this work we have computed the threshold resummation for Higgs boson pair production at hadron colliders via gluon fusion, including finite $M_t$ effects. We presented results both at NLL and NNLL accuracy, consistently matched to the corresponding fixed order cross sections.

Our NLL+NLO predictions retain the full $M_t$ dependence, and represent the most advanced prediction for this process computed in the full theory, i.e. not relying on the large-$M_t$ limit.
We found that at 13~TeV the NLL+NLO cross section is larger than the NLO result by about $4.1\%$ for the central scale $\mu_0 = M_{hh}/2$, while this effect goes up to $16.7\%$ for $\mu_0 = M_{hh}$.
The size of the resummed contributions decreases with the energy, going down to $2.8\%$ and $1.7\%$ at 27 and 100~TeV respectively, again for $\mu_0 = M_{hh}/2$.
We observed clear differences in the shape of the corrections as a function of $M_{hh}$ with respect to the large-$M_t$ result, but moderate compared to the overall size of the threshold effects.

Using the knowledge of the full NLL contributions, we have defined an improved NNLO approximation, \nnloFTi. We found that the difference with respect to the \nnloFT of Ref.~\cite{Grazzini:2018bsd} is very small, always below $0.5\%$ for all the collider energies under consideration and well within the estimated $M_t$ uncertainties of the approximation, pointing towards the reliability of the \nnloFT result.

Finally, we have also consistently combined our full NLL predictions with the NNLL resummation computed in the large-$M_t$ limit, and matched it to the \nnloFTi result, thus providing a prediction for the Higgs boson pair production cross section with the most advanced ingredients available to date.
We found that the effect of the resummed contributions is small at this order, being about $0.8\%$ at the LHC and smaller for larger collider energies. The effect is again larger for $\mu_0 = M_{hh}$, being around $8.1\%$ at 13~TeV.
The small size of the threshold resummation effects at NNLL, specially for $\mu_0 = M_{hh}/2$, is an indication of the fact that the perturbative expansion is under good control, and that no sizeable higher order effects are expected beyond the order reached within this calculation.

\subsection*{Acknowledgements}

We would like to thank Massimiliano Grazzini for valuable discussions and Gudrun Heinrich for helpful comments on the manuscript.
This research was supported in part by the Swiss National Science Foundation (SNF) under contracts CRSII2-141847, 200020-169041, by the 
Forschungskredit of the University of Zurich, by Conicet and ANPCyT.

\bibliography{biblio}

\end{document}